\documentclass[10pt,journal,a4paper,fleqn]{IEEEtran}
\usepackage{amsmath}
\usepackage{graphicx}
\usepackage{epstopdf}

\usepackage{amssymb}
\usepackage{caption}
\usepackage{subcaption}
\usepackage{bbm}
\usepackage{float}
\let\labelindent\relax
\usepackage{atbegshi}% http://ctan.org/pkg/atbegshi
\AtBeginDocument{\AtBeginShipoutNext{\AtBeginShipoutDiscard}}
\usepackage{enumitem}
\begin{document}
\title{Robust Nonparametric Sequential Distributed Spectrum Sensing under EMI and Fading}
\author{Sahasranand K. R. and Vinod Sharma,~\IEEEmembership{Senior Member, IEEE}}\thanks{This work was partly presented in IEEE National Conference on Communications, 2015 and IEEE International Conference on Communications, 2015. Partially supported by a grant from ANRC.}\thanks{The authors are with the department of Electrical Communication Engineering, Indian Institute of Science, Bangalore, India. Email: \{sanandkr, vinod\}@ece.iisc.ernet.in}\maketitle
\begin{abstract}
A nonparametric distributed sequential algorithm for quick detection of spectral holes in a Cognitive Radio set up is proposed. Two or more local nodes make decisions and inform the fusion centre (FC) over a reporting Multiple Access Channel (MAC), which then makes the final decision. The local nodes use energy detection and the FC uses mean detection in the presence of fading, heavy-tailed electromagnetic interference (EMI) and outliers. The statistics of the primary signal, channel gain or the EMI is not known. Different nonparametric sequential algorithms are compared to choose appropriate algorithms to be used at the local nodes and the FC. Modification of a recently developed random walk test is selected for the local nodes for energy detection as well as at the fusion centre for mean detection. It is shown via simulations and analysis that the nonparametric distributed algorithm developed performs well in the presence of fading, EMI and is robust to outliers. The algorithm is iterative in nature making the computation and storage requirements minimal.
\end{abstract}

\section{Introduction}
\label{section:intro}

Spectrum has been a costly commodity of late and intelligent use of available spectrum is warranted. A paradigm that helps us share the available spectrum is called Cognitive Radio (CR) \cite{mitola}. When the licensed users (primary users) are not using the spectrum, others (secondary users) can make use of it provided they sense the availability as quickly as possible. This problem is known as Spectrum Sensing in CR literature. Depending upon the knowledge of the primary signalling and the channel gains (\cite{CR_survey}, \cite{CR_poor_survey}), spectrum sensing is performed in a wide variety of ways.

\par There is a need to detect the presence of holes as early as possible to make efficient use of idle channel and to minimize interference to the primary users. Hence \emph{sequential} procedures serve better which can reduce the expected number of samples required, by more than half, over the fixed sample procedures \cite{govind}. Detection of spectral holes has to be performed at very low SNRs ($\sim$ --20 dB) in the presence of shadowing and fading \cite{low_snr_04}. This also demands \emph{distributed} detection which exploits spatial diversity to mitigate fading and can also reduce the detection time (\cite{CR_survey}, \cite{CR_poor_survey}). Furthermore, the transmit power, channel gains, coding and modulations of the primary are unknown and hence standard algorithms such as matched filter or cyclostationarity detector (\cite{CR_poor_survey}) may not be available. Energy detection (or generalised energy detection \cite{atapattu}) is found to be the technique applicable in such scenarios. Lack of complete knowledge about the signal and the channel fading (shadowing) calls for \emph{nonparametric (or semiparametric)} detection algorithms. Besides, the distribution of SINR may not be known and noise power could be time varying due to time varying electromagnetic interference (EMI). EMI is modelled using heavy-tailed distributions (\cite{SalphaS_survey}, \cite{SalphaS_trans}) and outliers \cite{huber} could be present in the samples received at the local nodes as well as the fusion centre (FC) over a reporting Multiple Access Channel (MAC). Channel fading can have Rayleigh, Rician or Nakagami distribution and shadowing is modelled by log normal distribution \cite{tse}, \cite{nakagami}. Thus the channel gain could possibly have a heavy-tailed component (due to log normal distribution) and a light-tailed component (due to the fading component) \cite{logray}. Hence \emph{robust} tests which work well with heavy tailed noise and signals are required. In summary, it is desirable to have distributed, nonparametric, robust, sequential algorithms for spectrum sensing in a CR system which mitigate the effects of heavy tailed distributions also.

\par Spectrum sensing has been subjected to detailed study during the recent years. \cite{mitola}, \cite{ghas}, \cite{viswanath} and the references therein give an overview of pioneering work in spectrum sensing. See \cite{CR_survey, CR_poor_survey, jointPHY, latest, hypt_survey} for more recent contributions. Various studies have suggested parametric (\cite{fellouris}, \cite{mei}) as well as nonparametric (\cite{uscslrt}, \cite{entropy_icc}) solutions to this problem. None of these works studies the effect of EMI or outliers on the detection algorithm. Distributed spectrum sensing has been a recent development in this direction (\cite{CR_survey}, \cite{chamberland, valli, taposh} and the references therein). See \cite{hypt_survey, jithin_ncc, quan, dualsprt, liusensor, jayakrishnan, banavar} for more recent developments in distributed detection and \cite{banavar} for distributed estimation. Some of the issues in distributed detection are that the reporting channel (for decisions from the local nodes to the FC) should not require much bandwidth and the energy consumed and the delay in reporting the decisions should also be small \cite{CR_survey}. Many of the works (\cite{CR_survey}, \cite{CR_poor_survey}, \cite{mei}, \cite{no_noiseFC}) do not consider MAC noise or multipath fading in the reporting channel. However, see \cite{multipath} and the references therein for studies which consider shadowing and fading in reporting channels. Design of algorithms at the local nodes as well as the fusion centre are motivated by the various above considerations.
\par The contribution of this paper is in designing new distributed, sequential, nonparametric energy detection and mean detection algorithms which perform well in the presence of slow-fast fading, heavy-tailed EMI and outliers. We are not aware of any other robust nonparametric scheme to mitigate the effects of EMI and outliers. Theoretic analysis of the algorithm is also provided.

\par The paper is organized as follows. Section~\ref{section:model} provides the system model and the distributed set up. Section~\ref{section:survey} presents several available (nonparametric) algorithms and their comparison via simulations. It also selects appropriate algorithms for the local nodes and FC for our distributed algorithm. Section~\ref{section:single node analysis} provides theoretical performance analysis of selected algorithms. Section~\ref{section:distributed analysis} theoretically analyses the distributed algorithm. It also shows the effect of heavy tails on the system performance. Section~\ref{section:approx} provides an approximation analysis of the algorithm. Section~\ref{section:distri simulation} provides the performance of the distributed algorithm for specific examples via simulations. Section~\ref{section:conclusion} concludes the paper.

\section{System Model and distributed algorithm}
\label{section:model}
%(Figure~\ref{fig:districog})
We consider a CR system where $L$ CR (local) nodes are scanning the environment to detect if a primary user is transmitting or not. Based on their observations, the nodes make local decisions and transmit to the FC. The FC makes the final decision based on the local decisions it receives from the secondary nodes. This is the most common distributed spectrum sensing architecture (\cite{CR_survey}, \cite{CR_poor_survey}).

\par At time $k$, node $l$ senses $\tilde{X}_{kl}$ (at baseband level) where
\begin{equation*}
\tilde{X}_{kl} = H_{kl}S_{k} + N_{kl}
\end{equation*}
if a primary is transmitting (Hypothesis $\mathcal{H}_1$). Here, at time $k$, $H_{kl}$ is the channel gain from the primary to the local node $l$, $S_k$ is the symbol transmitted by the primary and $N_{kl}$ is the node $l$ receiver noise with possibly some EMI. If the primary is not transmitting at time $k$ (Hypothesis $\mathcal{H}_0$) then
\begin{equation*}
\tilde{X}_{kl} = N_{kl}.
\end{equation*}
We assume that $\{S_{k}, k \ge 1\}$ and $\{N_{kl}, k \ge 1\}$ are independent identically distributed (i.i.d.) and independent of each other. In the following this assumption will be slightly generalized. Also, $\{N_{kl}\}$ are assumed independent sequences for different nodes $l$.

\par For $\{H_{kl}, k \ge 1\}$, we either assume that $H_{kl} \equiv H_{l}$, a random variable, possibly unknown (this is a commonly made assumption \cite{tse}, \cite{simon_alouni}), representing slow fading, or an i.i.d. sequence, representing fast fading. $H_{kl}$ represents multipath fading as well as shadowing. For shadowing, log normal distribution is considered a good approximation \cite{tse}, while for multipath fading, Rayleigh, Rician and Nakagami distributions are considered suitable \cite{nakagami}. Thus $H_{kl}$ could possibly have a heavy-tailed component (due to log normal distribution) and a light-tailed component (due to the fast fading component) \cite{logray}. Often the combined effect of these is approximated by a K-distribution \cite{K_distri} which has a heavy tail.

\par If sensing is done at times of primary symbol transmission then assuming $\{S_k\}$ to be i.i.d. is realistic which will often take values in a finite alphabet depending on the modulation scheme used by the primary. The secondary may not know the coding and modulation used by the primary. Also, different primary users may be using the same channel and a primary can change its modulation and coding with time. Thus, we will not assume that the local nodes know the signalling of the primary. This is a common assumption in the CR literature.

\par As a result of unknown $H_{kl}, S_{k}$ statistics, it is usually recommended to use energy detection at the local nodes (\cite{CR_survey}, \cite{CR_poor_survey}). Thus, we consider the energy samples
\begin{equation}
X_{kl} = \sum_{i=(k-1)M+1}^{Mk} {(\tilde{X}_{il})}^2
\label{equation:energy}
\end{equation}
at each local node $l$ where $M$ is a constant decided as part of the sensing algorithm. Taking square of $\tilde{X}_{kl}$ in (\hspace{-0.12cm}~\ref{equation:energy}) provides the usual energy detector and is shown to be optimal for Gaussian noise in the absence of $S_k$ statistics. However, it has been shown \cite{atapattu} that for non Gaussian noise, instead of $2$, some other power $p$ of $|\tilde{X}_{kl}|$ may perform better. In the following we will keep $p=2$ but allow the possibility of other powers when EMI is significant (see below).

\par In the following we will only assume $\{X_{kl}, k \ge 1\}$ to be i.i.d. independent sequences under $\mathcal{H}_0$ and $\mathcal{H}_1$ allowing $\{\tilde{X}_{il}, Mk+1 \le i \le M(k+1)\}$ to have arbitrary dependence. This provides flexibility in modelling fading and sensing versus signalling duration. 

\par The receiver noise is usually distributed as Gaussian, mean $0$ and variance (say) $\sigma^2$ (denoted as $\mathcal{N}(0,\sigma^2)$). However, in wireless channels there can often be a significant component of EMI \cite{SalphaS_survey}. EMI is modelled by Gaussian mixtures (which are light-tailed) and symmetric $\alpha$-stable distributions (which are heavy-tailed for $\alpha<2$) (\cite{SalphaS_trans}). Thus $N_{kl}$ will often not be Gaussian and can possibly be heavy-tailed. Of course, as a result of squaring $\tilde{X}_{kl}$, the noise distribution will not be symmetric.

\par Now we consider the hypothesis testing problem one encounters for energy detection with samples (\hspace{-0.12cm}~\ref{equation:energy}). We will denote by $P_i$, $\mathbb{E}_i[X]$ and $Var_i[X]$, the distribution, the mean and the variance of $X$ under the hypothesis $\mathcal{H}_i$, $i=0,1$. For simplicity, we take $\{\tilde{X}_{kl}, k \ge 1\}$ i.i.d. in this paragraph. If $N_{kl}$ has a general distribution with mean 0 and variance ${\sigma_l}^2$, under $\mathcal{H}_0$, $\mathbb{E}_0[X_{1l}] = M{\sigma_l}^2$ and $Var_0(X_{1l}) = M(\mathbb{E}_0[(N_{1l})^4] - {\sigma_l}^4)$. Also, under $\mathcal{H}_1$, $\mathbb{E}_1[X_{1l}] = M{\sigma_{1l}}^2 + E_{sl}$ and $Var_1(X_{1l}) = M(\mathbb{E}_1[(N_{1l} + H_{1l}S_{1l})^4] - ({\sigma_{1l}}^2 + \mathbb{E}_1[(H_{1l}S_{1l})^2])^2)$ where $E_{sl} = M\mathbb{E}_1[(H_{1l}S_{1l})^2]$, the received energy at node $l$.

\par If ${\sigma_l}^2 >> E_{sl}$ and ${\sigma_l}^2, E_{sl} $ are known but the distributions of $N_{kl}, S_k$ are not known, we can consider it as a nonparametric mean detection problem with $\mathcal{H}_0:\mu=\mu_0=M{\sigma_l}^2$ vs $\mathcal{H}_1:\mu=\mu_1=M{\sigma_l}^2 + E_{sl}$. It is a simple hypothesis testing problem with equal known variance under both hypotheses. If $E_{sl}$ is not known but we know that $E_{sl}$ is lower bounded by $E_L$ then the testing problem is $\mathcal{H}_0:\mu=\mu_0=M{\sigma_l}^2$ vs $\mathcal{H}_1:\mu = M{\sigma_l}^2 + E_s \ge M{\sigma_l}^2 + E_L = \mu_1$. Now $\mathcal{H}_1$ is a composite hypothesis. If ${\sigma_l}^2$ is also not known but we know that ${\sigma_L}^2 < {\sigma_l}^2 < {\sigma_U}^2$ then the problem is $\mathcal{H}_0:\mu = M{\sigma_l}^2 \le M{\sigma_U}^2 = \mu_0$ and $\mathcal{H}_1:\mu = M{\sigma_l}^2 + E_s \ge M{\sigma_L}^2 + E_L = \mu_1$. Now the variance under the two hypotheses are the same but unknown. The most general situation arises when the low SNR assumption is also violated and now the unknown variances under the two composite hypotheses are not the same.

\par As a consequence of the above comments, for a local node to make a decision, nonparametric statistical techniques which do not require complete knowledge of the distributions of observations $X_{kl}$ under $\mathcal{H}_0$ and $\mathcal{H}_1$ are suitable for energy detection. To make quick decisions, local nodes will use sequential detection. Thus node $l$ will make its decision at a random time based on its local observations $\{X_{kl}, k \ge 1\}$. In the next section we compare several nonparametric sequential algorithms for energy detection and pick the best.

\par If node $l$ decides $\mathcal{H}_1$ at time $k$, it will transmit $+b_1$ to the FC. If it decides $\mathcal{H}_0$, it transmits $-b_0$. If the node has not made a decision at a time, it transmits nothing. Thus, at time $k$, FC receives $Y_k = \displaystyle \sum_{l=1}^{L} G_{kl}Y_{kl} + Z_{k}$ where $Y_{kl}$ is the transmission from node $l$, $G_{kl}$ is the corresponding channel gain and $Z_k$ is the superposition of the receiver noise (which will often have a distribution $\mathcal{N}(0,\sigma^2)$) and EMI. Thus, $Z_k$ will be a summation of Gaussian noise and Gaussian mixtures and/or alpha-stable EMI. The distribution of $G_{kl}$ may also not be known. Thus, we need at the FC a nonparametric sequential algorithm but unlike at the local nodes, the signalling ($+b_1$ or $-b_0$) is known to the FC. Furthermore, unlike at the local nodes, we can use partially coherent detection (we may be able to estimate the phase; in particular, the sign of $G_{kl}$ although not necessarily the magnitude of the channel gains \cite{partial_coherent}, \cite{tse}). Then the local node multiplies its transmission $Y_{kl}$ ($+b_1$ or $-b_0$) by the sign of $G_{kl}$ and transmits. Thus, $Y_k = \displaystyle \sum_{l=1}^{L} |G_{kl}|Y_{kl} + Z_{k}$. Therefore we do not need an energy detector (actually in our set up we may not be able to use the energy detector at the FC) but in fact a nonparametric detector which performs well for mean detection with symmetric noise will be a suitable choice (if $Z_k$ is zero mean symmetric, which will often happen in practice. But we will not assume symmetric distribution in the following). 

\par As discussed above, at the local nodes as well as at the FC, due to possibly significant EMI, the noise may be heavy-tailed. Such a scenario in CR has been considered in \cite{SalphaS_survey}. But the impact of heavy-tailed noise has not been specifically studied. In \cite{taposh}, this was considered in the context of change detection and it was shown that heavy tails can degrade the performance significantly. In this paper, for the distributed hypothesis testing algorithm also, we show that heavy-tailed distributions can significantly impact the performance. Then we will modify the algorithms so that their impact along with that of the outliers which are also present, can be mitigated.

\par Often the reporting (MAC) channel from the local nodes to the FC is considered noiseless (\cite{CR_survey}, \cite{mei}, \cite{CR_poor_survey}, \cite{no_noiseFC}). However, as mentioned above, like any other wireless channel, it does experience EMI, outliers and receiver noise. One implication of this is that the decisions transmitted by local nodes may not reach the FC without error making the use of standard Fusion centre rules - AND, OR, majority etc. \cite{CR_poor_survey} less accurate and/or difficult to implement.

\par Now we describe our basic distributed algorithm which has been shown to be asymptotically optimal and performs well at practical parameter values (\cite{dualsprt}, \cite{entropy_icc}). It also makes an efficient use of the reporting MAC. An optimal algorithm in this setting is not known \cite{valli}. We will complete this algorithm by choosing appropriate detection algorithms for the local nodes and the FC in the next sections. We will also study the performance of the overall algorithm so developed especially under the influence of EMI, outliers and fading.

\subsection*{Distributed Algorithm}
\label{section:distri}

\begin{itemize}
\item Each local node $l$ receives observation $X_{kl}$ at time $k$.
\item Each node $l$ uses a sequential algorithm to compute $T_{kl} = f(X_{k,l},X_{(k-1),l}, ..., X_{1,l})$ and makes a decision at time $N_l$ where 
\begin{equation*}
N_l = inf\{n: T_{nl} \notin (-\gamma_{0l}, \gamma_{1l})\},
\end{equation*}
$\gamma_{0l}, \gamma_{1l}$ are appropriately chosen positive constants and the decision is $H_0$ if $T_{N_ll} \le -\gamma_{0l}$ and $H_1$ if $T_{N_ll} \ge \gamma_{1l}$. It transmits $Y_{kl}$ to the FC at time $k$ where
\begin{equation*}
Y_{kl} = b_1\mathbbm{1}\{T_{kl} \ge \gamma_{1l}\} - b_0\mathbbm{1}\{T_{kl} \le -\gamma_{0l}\}.
\end{equation*}
Node $l$ will keep transmitting till the FC makes a decision.
\item At time $k$, FC receives 
\begin{equation*}
Y_k = \displaystyle \sum_{l=1}^{L} Y_{k,l} + Z_{k}
\end{equation*}
and computes $W_k$ based on an algorithm to be decided. At time
\begin{equation*}
N= inf\{n: W_n \notin (-\beta_0, \beta_1))\},
\end{equation*}
it decides $H_1$ if $W_N \ge \beta_1$ and $H_0$ if $W_N \le -\beta_0$ where $\beta_0, \beta_1$ are appropriately specified. After $N$, all nodes stop transmitting. \hfill $\square$
\end{itemize}
\par The energy detection algorithm to be used by the local nodes and the mean detection to be used at the FC will be chosen in the next section.

\par One of the advantages of our distributed algorithm is that the local node $l$ which has a good channel gain $H_{kl}$ from the primary will make a decision faster and will influence the FC decision more. Also, since each local node keeps transmitting its decision till the FC decides, if a local node has made a wrong decision, most likely it will soon change it and hence wrong local decisions will have minimal effect on the FC decision, especially when $P_{FA}$ (probability that the FC decides $\mathcal{H}_1$ while $\mathcal{H}_0$ is true) and $P_{MD}$ (probability that the FC decides $\mathcal{H}_0$ while $\mathcal{H}_1$ is true) are small. 

\section{Single node: Algorithms}
\label{section:survey}
In this section we consider sequential nonparametric single node algorithms with their statistics denoted by $T_n$, which can be used by the local nodes and the FC for energy detection and mean detection respectively. Optimal tests for single nodes also do not exist. We will not use the node index $l$ in this section.
\subsection{Rank test}
Rank test (Wilcoxon rank test) is a location test \cite{govind} for location $\mu$ of a distribution $F(x-\mu)$ which is symmetric around $\mu$. For testing $\mu \le \mu_0$ vs $\mu \ge \mu_1, \mu_1 > \mu_0$, its statistics is defined as follows.
\begin{enumerate}
\item[i.] Let $Y_i = X_i - \frac{\mu_0 + \mu_1}{2}$, where $X_i$s are the observations.
\item[ii.] Calculate $R_i$, the rank of $Y_i$ in $Y_1, ..., Y_n$ when these are arranged in ascending order of their absolute values.
\item[iii.] Test statistic $T_n = \sum_{i=1}^{n} sgn(Y_i)\frac{R_i}{n+1}$ where $sgn(x)=\frac{x}{|x|}$ for $x \neq 0$ and $0$ for $x=0$.
\end{enumerate}
We will use this statistic in our sequential set up. This statistic is distribution free for symmetric distributions \cite{govind}.
\subsection{Sequential $t$ test}
We use the usual $t$ test \cite{lehmann} extended to make it a two sided test. The test statistic is given by, \\
\begin{equation}
T_n = n\frac{\overline{X}_n-\frac{\mu_0 + \mu_1}{2}}{s_n}
\end{equation}
where $\overline{X}_n = \frac{1}{n}\sum_{k=1}^{n} X_{k}$ is the sample mean,  and $s_n = [\frac{1}{n-1}\sum_{k=1}^{n} (X_{k}-\overline{X}_n)^2]^{1/2}$ is the sample variance.\\

\subsection{Random walk}
Its test statistic is obtained by modifying the above $t$ test statistic:
\begin{equation}
T_n = \sum_{i=1}^{n}(X_i-\frac{\mu_0 + \mu_1}{2}).
\label{equation:random}
\end{equation}
The statistic is a simple random walk and we refer to this algorithm as random walk.\\
\par The above three tests are primarily designed for detection of mean $\mathcal{H}_0: \mu \le \mu_0$ vs $\mathcal{H}_1: \mu \ge \mu_1$, but can also be used for testing some other functional of the distributions. Unlike sequential $t$ test and rank test, random walk test is \emph{iterative}. Thus it is simpler to compute the statistic and does not require storing the whole data.

\subsection{Mitigating effects of outliers, heavy tails and fading}
\label{subsection:robust}
The sample mean and the sample variance used in the $t$ test and random walk are not robust to outliers. This gets reflected in the performance of these tests (compare Figures~\ref{fig:gaussian_alpha} and~\ref{fig:gaussian_alpha_outlier} below; see also Figure~\ref{fig:compare_randomwalk}). From Figures~\ref{fig:gaussian_alpha} and~\ref{fig:gaussian_alpha_outlier} we also see that the rank test is quite robust to outliers although may not perform the best.  This motivates the use of robust versions of the random walk and $t$ tests \cite{huber}. Robust tests are obtained by replacing the sample mean (and sample variance) in these tests by their robust versions.
\begin{itemize}
\item $M-t$ test is obtained by applying a cut-off function $\psi$ (called \emph{Huber function} after \cite{huber}) to obtain a robust sample mean (corresponding modified sample variance is in the denominator of $T_n$ below.) and obtain the statistics of $t$ test as\\
\begin{equation}
T_{n} = \frac{\sum_{i=1}^{n}\psi(X_{i}-\frac{\mu_{0} + \mu_{1}}{2})}{(\sum_{i=1}^{n}\psi^2(X_{i}-\bar{X}_{n}))^\frac{1}{2}},
\label{equation:M t}
\end{equation} 
where $\psi : \mathcal{R} \mapsto  \mathcal{R} $ is a non decreasing, continuous, odd and bounded function. For $\mathcal{N}(0,1)$, a recommended $\psi$ \cite{huber} is
\begin{equation}
\psi_{0}(z) = 
\begin{cases}
K, \text{ if } z > K, \\
z, \text{ if } |z| \leq K, \\
-K, \text{ if } z < -K,
\end{cases}
\label{equation:huber}
\end{equation}

for a given positive $K < \infty$.
\item Applying the $\psi$ function on the random walk, we get a robust version called $M$-random walk via the statistic
\begin{equation}
T_{n} = \sum_{i=1}^{n}\psi(X_{i}-\frac{\mu_{0} + \mu_{1}}{2}).
\label{equation:M random}
\end{equation}
\par This statistic is \emph{iterative}, unlike the $t$ test or $M-t$ test.
\end{itemize}
\par It is known that the $t$ test is not efficient for heavy-tailed distributions \cite{lehmann}. One expects this behaviour for the random walk test also (see Figure~\ref{fig:compare_randomwalk} below). On the other hand, the rank test is quite efficient for heavy tailed distributions also.

\par We will also see that the Huber function $\psi$ not only robustifies $t$ and random walk tests but also makes them more efficient with respect to (w.r.t.) heavy-tailed distributions. We will confirm these findings from simulations and the theory in Section~\ref{section:single node analysis}.

\par In very heavy-tailed case (S$\alpha$S with $\alpha < 1$ or for energy detection with $\alpha < 2$), the mean of the sample $X_k$ is infinity. Thus, random walk and $t$ test will not work. The rank test can possibly still work. Even the above robust versions of random walk and $t$ test (\hspace{-0.12cm}~\ref{equation:M random}) and (\hspace{-0.12cm}~\ref{equation:M t}) will not work directly because $\mu_0$ and $\mu_1$ will be infinity. Thus, we replace samples $X_i$ with
\begin{equation}
\hat{X}_i = \psi_1(X_i)
\label{equation:M^2}
\end{equation}
where $\psi_1$ is from the class of functions mentioned below equation (\hspace{-0.12cm}~\ref{equation:M t}),
and use $M$-random walk test on it with $\mu_0$ and $\mu_1$ corresponding to the means of $\hat{X}_i$. We call this \emph{$M^2$-random walk} test. We will see below via simulations that $M^2$-random walk test works for S$\alpha$S with $\alpha < 2$ while $M$-random walk, random walk, $t$, $M-t$ and $M-t$ based on samples (\hspace{-0.12cm}~\ref{equation:M^2}) do not work at all.
\par Choice of $\psi$ in (\hspace{-0.12cm}~\ref{equation:M t}), (\hspace{-0.12cm}~\ref{equation:M random}) and $\psi_1$ in (\hspace{-0.12cm}~\ref{equation:M^2}) affects the performance of the algorithm (see \cite{huber} for different $\psi$ in parametric set up). In our nonparametric setup we will simply use $\psi_0$ defined in (\hspace{-0.12cm}~\ref{equation:huber}) with different $K$ values. Our aim of using $\psi$ for heavy-tailed case is to create light-tailed samples (\hspace{-0.12cm}~\ref{equation:M^2}). In our simulations below for energy samples, we will take $K$ large for $\psi_1$($\approx 200$) but small ($\le 5$) for $\psi$ in (\hspace{-0.12cm}~\ref{equation:M t}) and (\hspace{-0.12cm}~\ref{equation:M random}).

\par It has been known that slow fading can significantly degrade the performance of a detection algorithm (\cite{simon_alouni}). We will see that this happens for the above algorithms also. This is because in slow fading, $X_k =HS_k+N_k$ and for usual fading distributions e.g., Rayleigh, $H$ can be small with a large probability. In this case, applying the $\psi$ function does not help. Then if we do not make a decision when $|H| \le \delta$ for a small $\delta$, it can significantly improve the performance if we take $\mathbb{E}_H[\mathbb{E}_i[N(H)]$ as the performance measure for given $P_{FA}$ and $P_{MD}$ where $\mathbb{E}_i[N(H)]$ is the mean number of samples needed to decide under $\mathcal{H}_i$ when the channel gain is $H$. The constant $\delta$ needs to be chosen carefully depending on the desired probabilities of error. In the distributed setting, due to spatial diversity, the $\delta$ needed can be reduced. We will study the effect of this operation via simulation and theory in the following.
\par When both EMI and slow fading are present, then we should combine the above two operations: not make a decision if $|H| \le \delta$ and when we do make, we use (\hspace{-0.12cm}~\ref{equation:random}), (\hspace{-0.12cm}~\ref{equation:M random}) and (\hspace{-0.12cm}~\ref{equation:M^2}). We will call the corresponding random walk algorithms, $\delta$-random walk, $M-\delta$-random walk and $M^2-\delta$-random walk. Similarly we name the $t$ test.

\subsection{Simulation Results}
\label{subsection:simu}
\par We compare the above algorithms for mean detection when the channels may experience slow/fast fading with shadowing and S$\alpha$S EMI and outliers. This scenario can be useful for energy detection at low SNR and at the FC. We have taken $\alpha = 1.8$ for the S$\alpha$S distribution \cite{SalphaS_trans} and fading is Rayleigh distributed with parameter $P$ where $P \sim \log \mathcal{N}(0,0.36)$ represents shadowing \cite{logray}. The receiver noise $Z \sim \mathcal{N}(0,\sigma^2)$ and $SNR = 10\log\frac{\mathbb{E}[H^2]{(\mu_1-\mu_0)}^2}{\sigma^2}$. The $X$-axis shows $\frac{P_{FA}+P_{MD}}{2}$ and the $Y$-axis shows $\frac{\mathbb{E}_0[N]+ \mathbb{E}_1[N]}{2}$. For slow fading we keep the channel gains constant till the decisions are made. The simulations were run $10,000$ times and averaged to obtain the probabilities of error and the mean time to sense.

\begin{figure}[ht]
        \centering
                \includegraphics[width=0.45\textwidth,height=5cm]{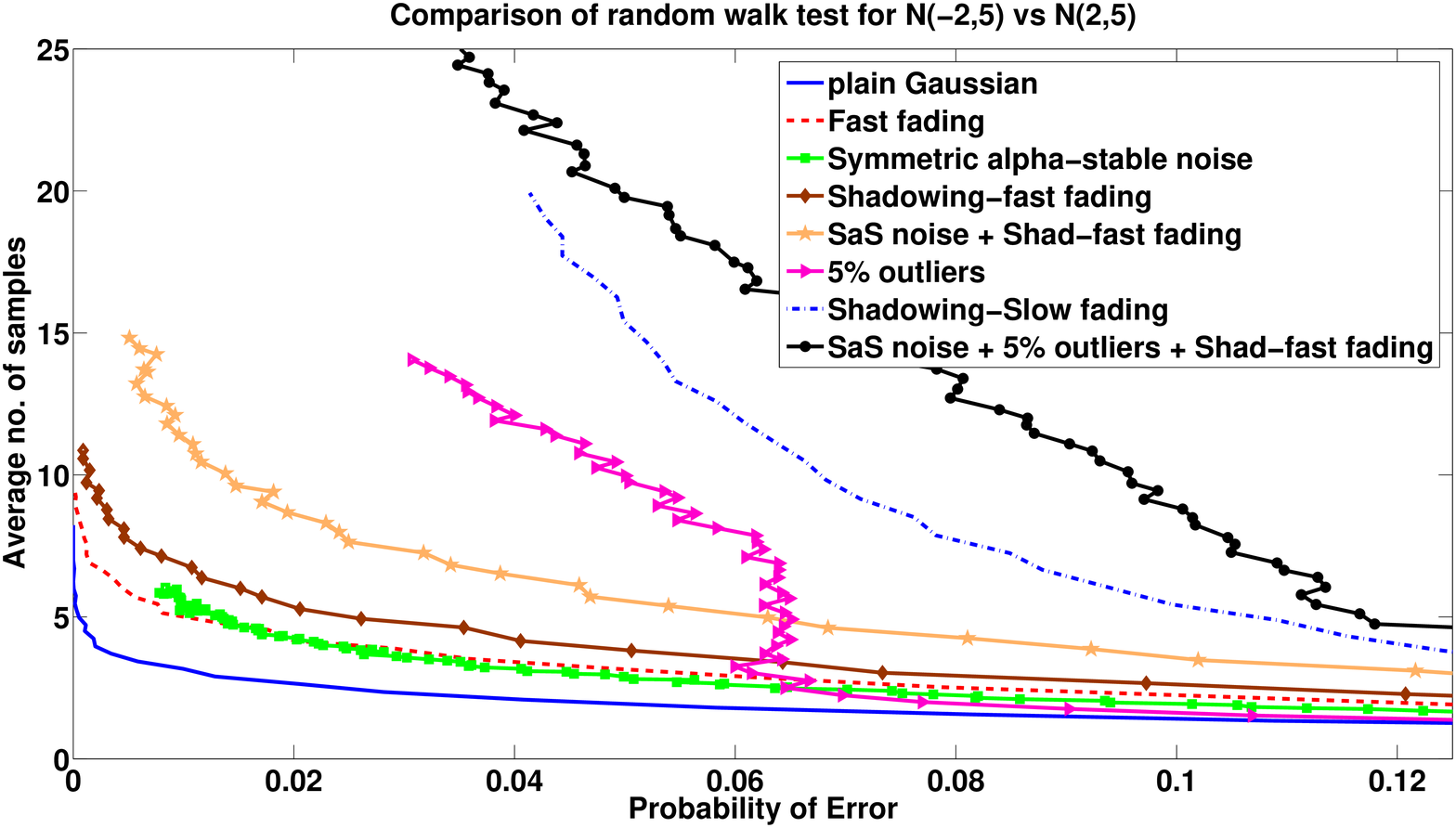}
\caption{\small Effect of different factors on the performance of random walk.}\label{fig:compare_randomwalk}
\end{figure}

\begin{figure}[]
        \centering
        \begin{subfigure}[b]{0.5\textwidth}
                \centering
                \includegraphics[width=\textwidth,height=5cm]{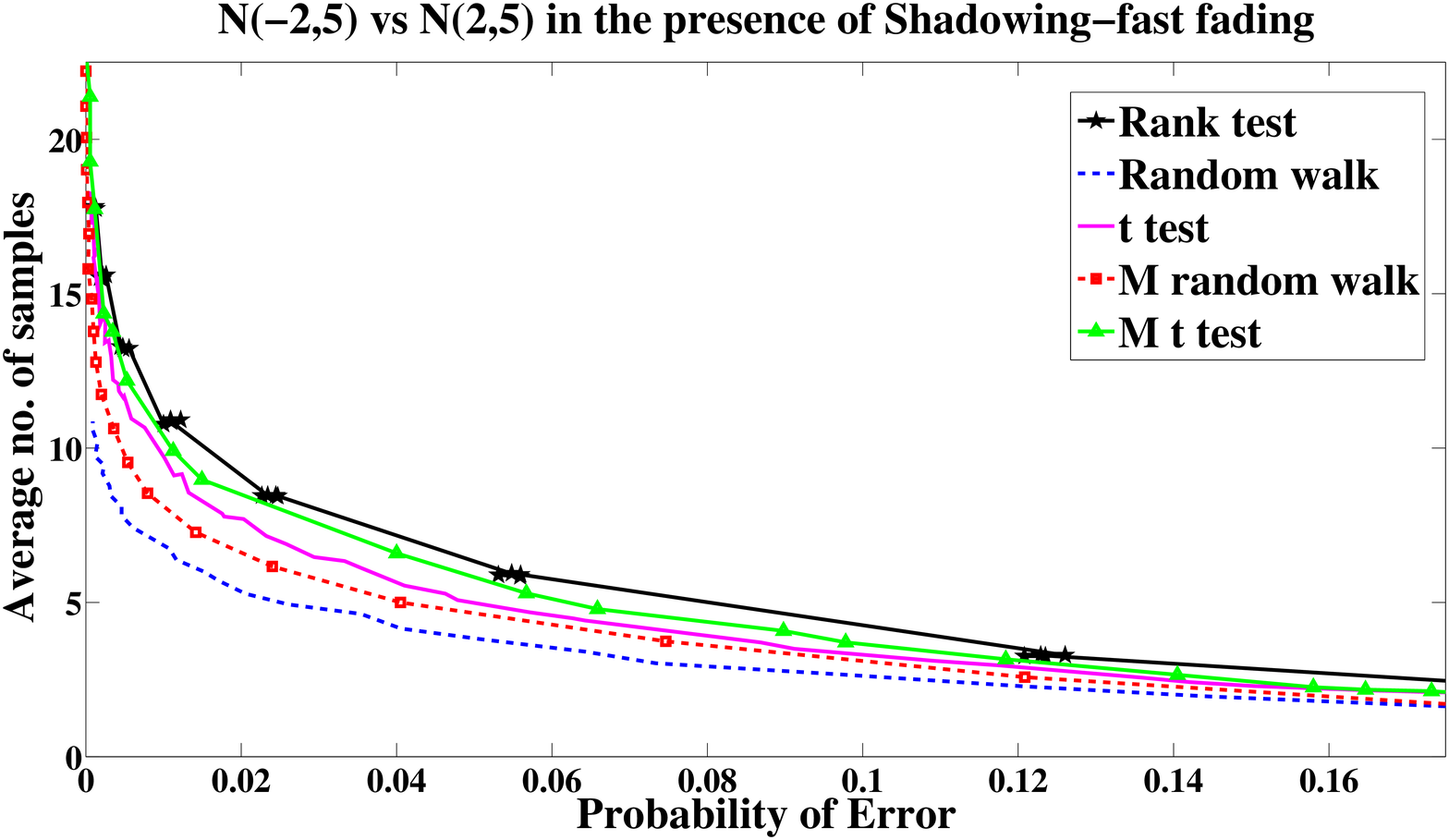}
                \label{fig:G_Sh_Ff}
        \end{subfigure}
        \begin{subfigure}[b]{0.5\textwidth}
                \centering
                \includegraphics[width=\textwidth,height=5cm]{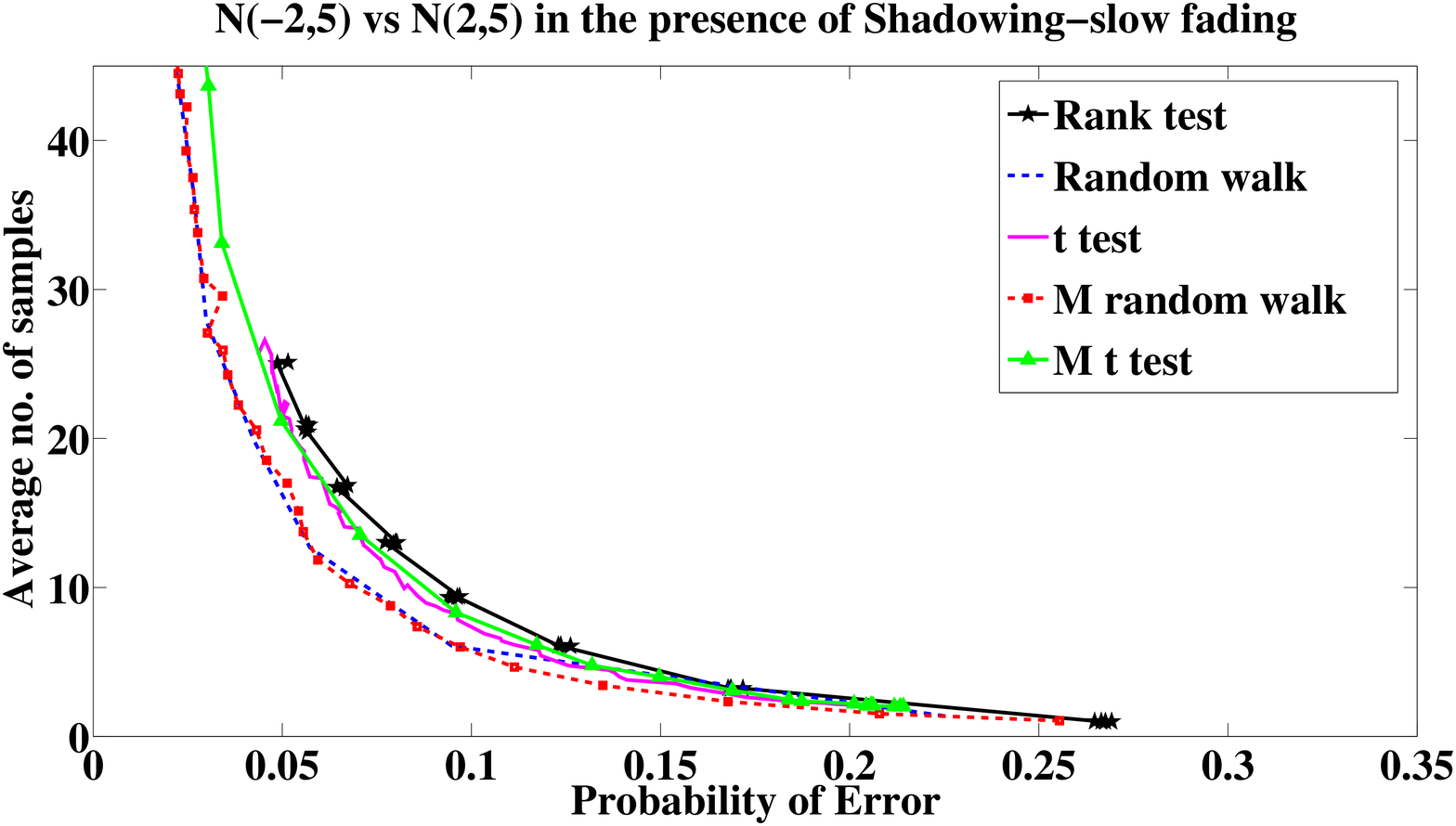}
                \label{fig:G_Sh_Sf}
        \end{subfigure}
        \caption{\small Mean detection at FC in the presence of Gaussian noise. Top: Log $\mathcal{N}$ shadowing - Rayleigh \emph{fast} fading. Bottom: Log $\mathcal{N}$ shadowing - Rayleigh \emph{slow} fading.}
        \label{fig:only_gaussian}
\end{figure}

\begin{figure}[h]
        \centering
        \begin{subfigure}[b]{0.5\textwidth}
                \centering
                \includegraphics[width=\textwidth,height=5cm]{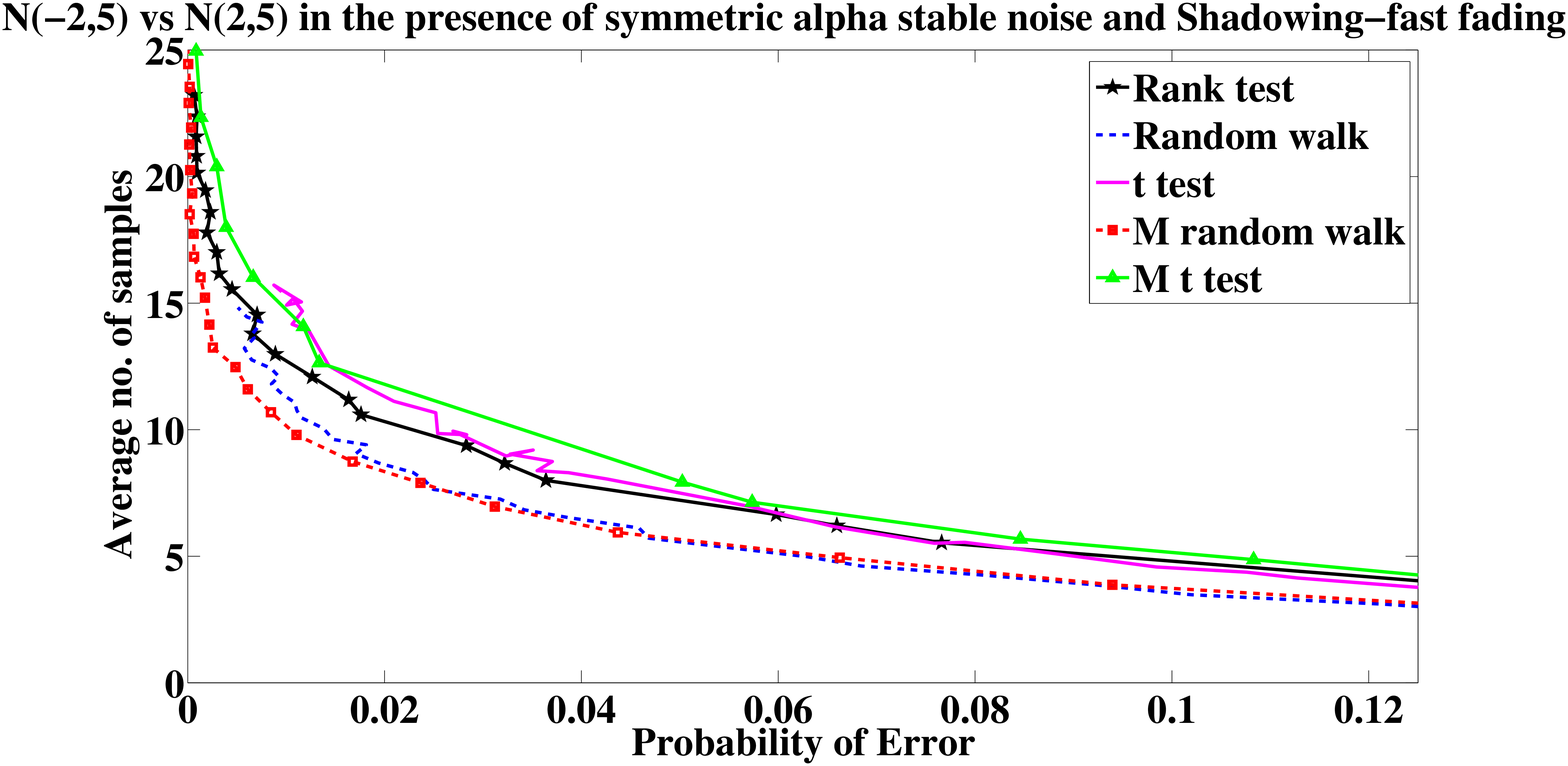}
                \label{fig:GSaS_Sh_Ff}
        \end{subfigure}
        \begin{subfigure}[b]{0.5\textwidth}
                \centering
                \includegraphics[width=\textwidth,height=5cm]{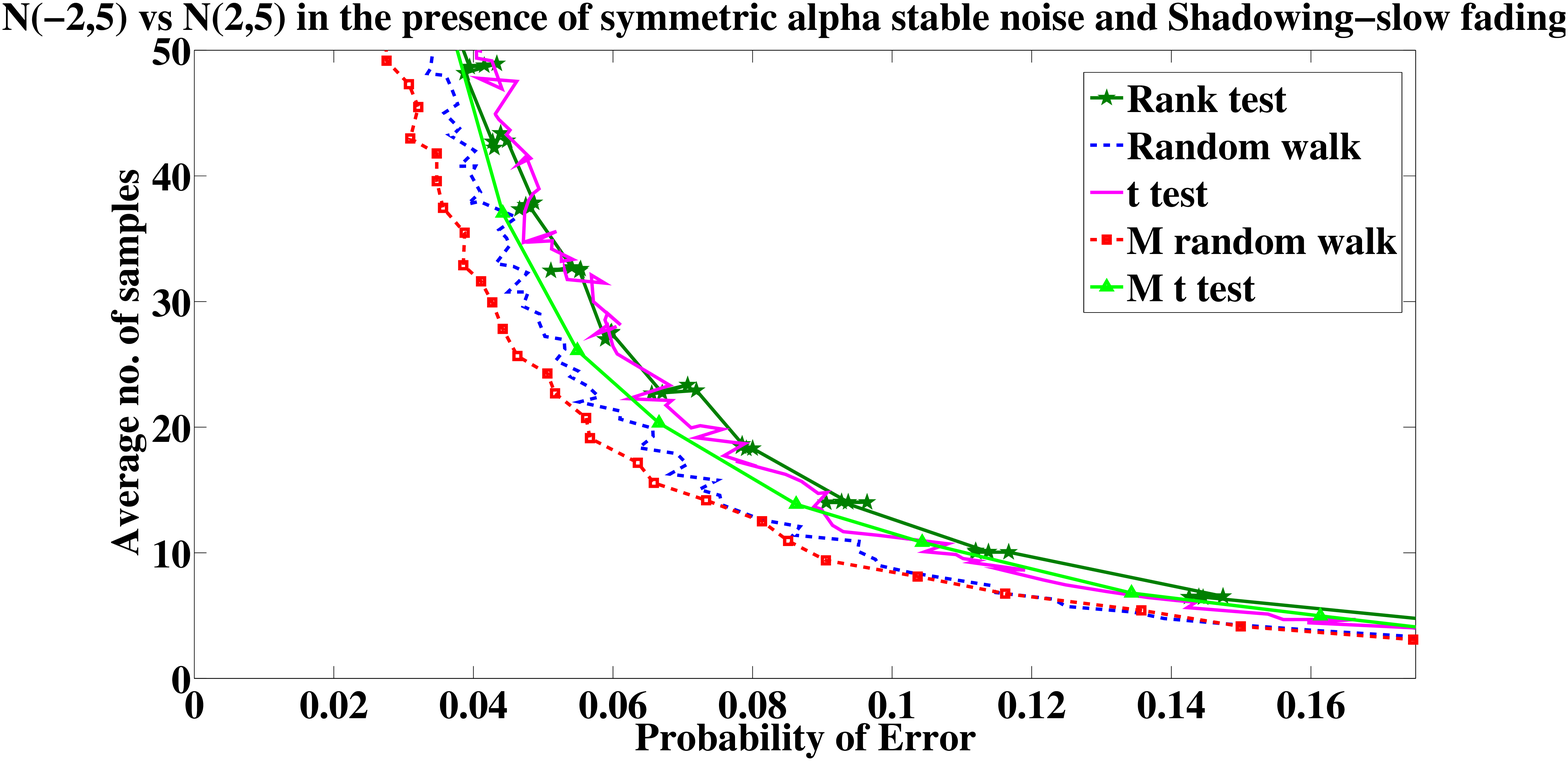}
                \label{fig:GSaS_Sh_Sf}
        \end{subfigure}
        \caption{\small Mean detection at FC in the presence of Gaussian and symmetric $\alpha$-stable noise. Top: Log $\mathcal{N}$ shadowing - Rayleigh \emph{fast} fading. Bottom: Log $\mathcal{N}$ shadowing - Rayleigh \emph{slow} fading.}
\label{fig:gaussian_alpha}
\end{figure}

\begin{figure}[]
        \centering
        \begin{subfigure}[b]{0.5\textwidth}
                \centering
                \includegraphics[width=\textwidth,height=5cm]{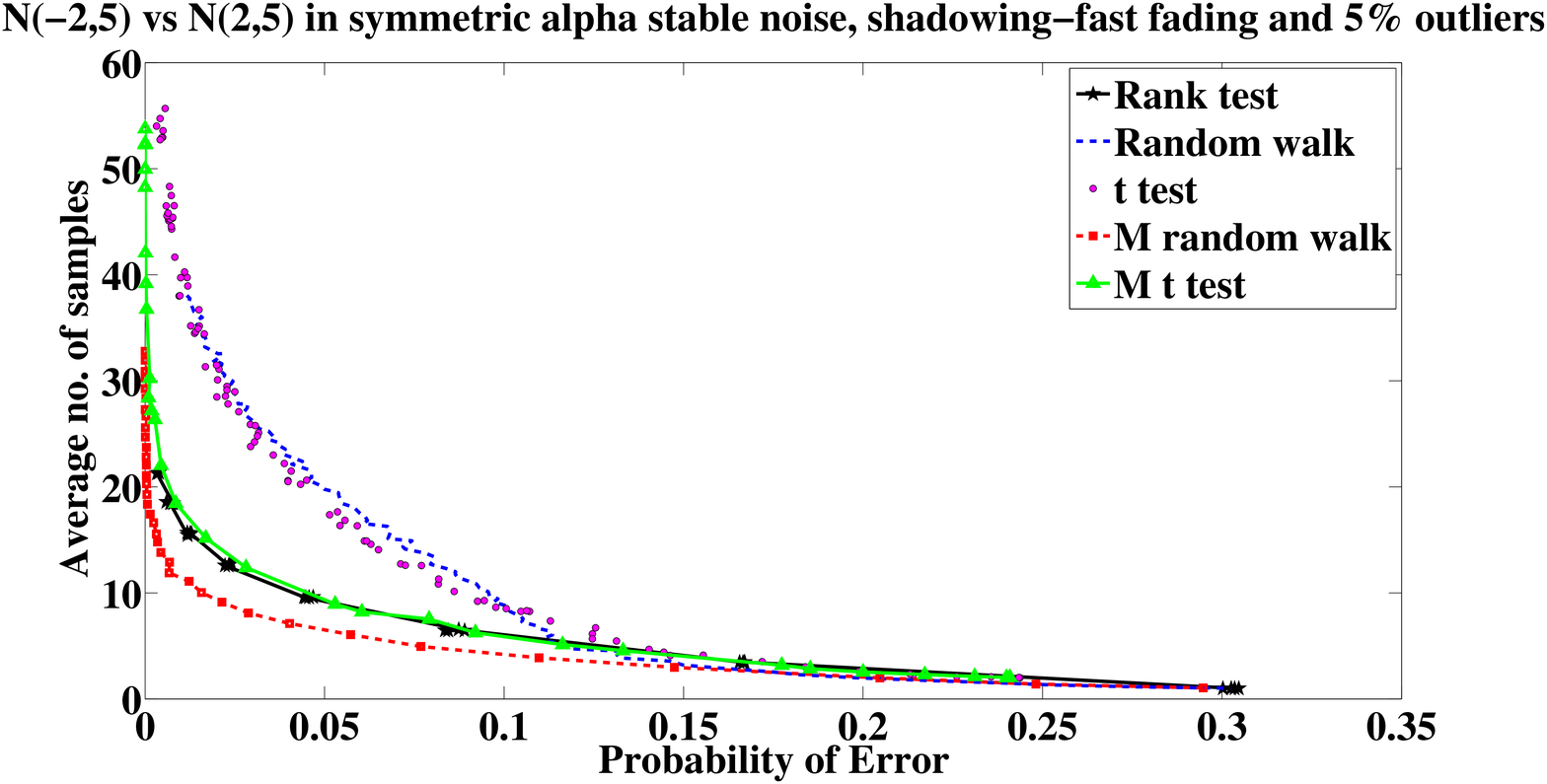}
                \label{fig:GSaSO_Sh_Ff}
        \end{subfigure}
        \begin{subfigure}[b]{0.5\textwidth}
                \centering
                \includegraphics[width=\textwidth,height=5cm]{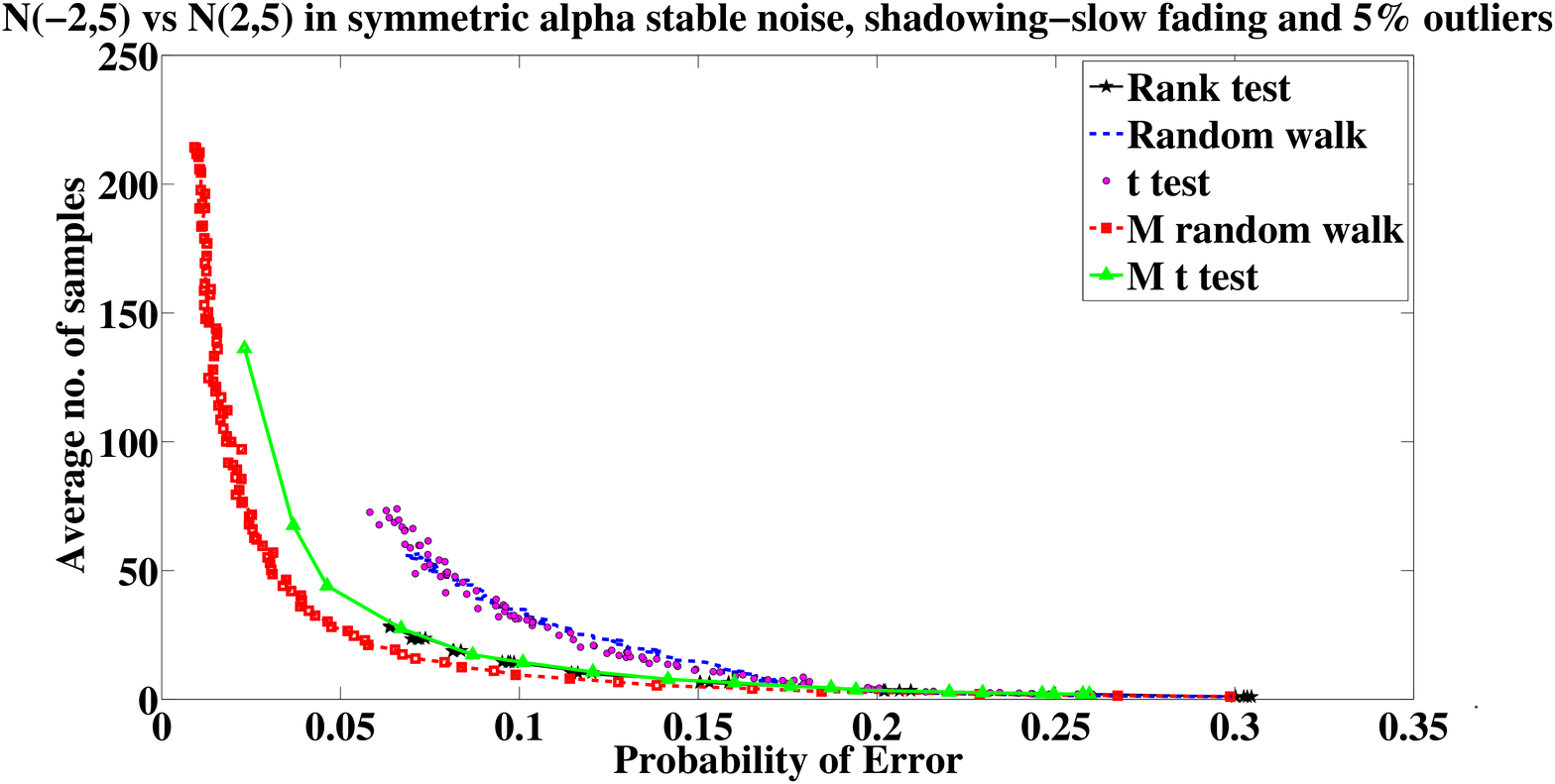}
                \label{fig:GSaSO_Sh_Sf}
        \end{subfigure}
         \caption{\small Mean detection at FC in the presence of Gaussian and symmetric $\alpha$-stable noise and $5\%$  $\mathcal{N}(0,20)$  \emph{outliers}. Top: Log $\mathcal{N}$ shadowing - Rayleigh \emph{fast} fading. Bottom: Log $\mathcal{N}$ shadowing - Rayleigh \emph{slow} fading.}
\label{fig:gaussian_alpha_outlier}
\end{figure}

\begin{figure}[h]
        \centering
        \begin{subfigure}[b]{0.5\textwidth}
                \includegraphics[width=\textwidth,height=5cm]{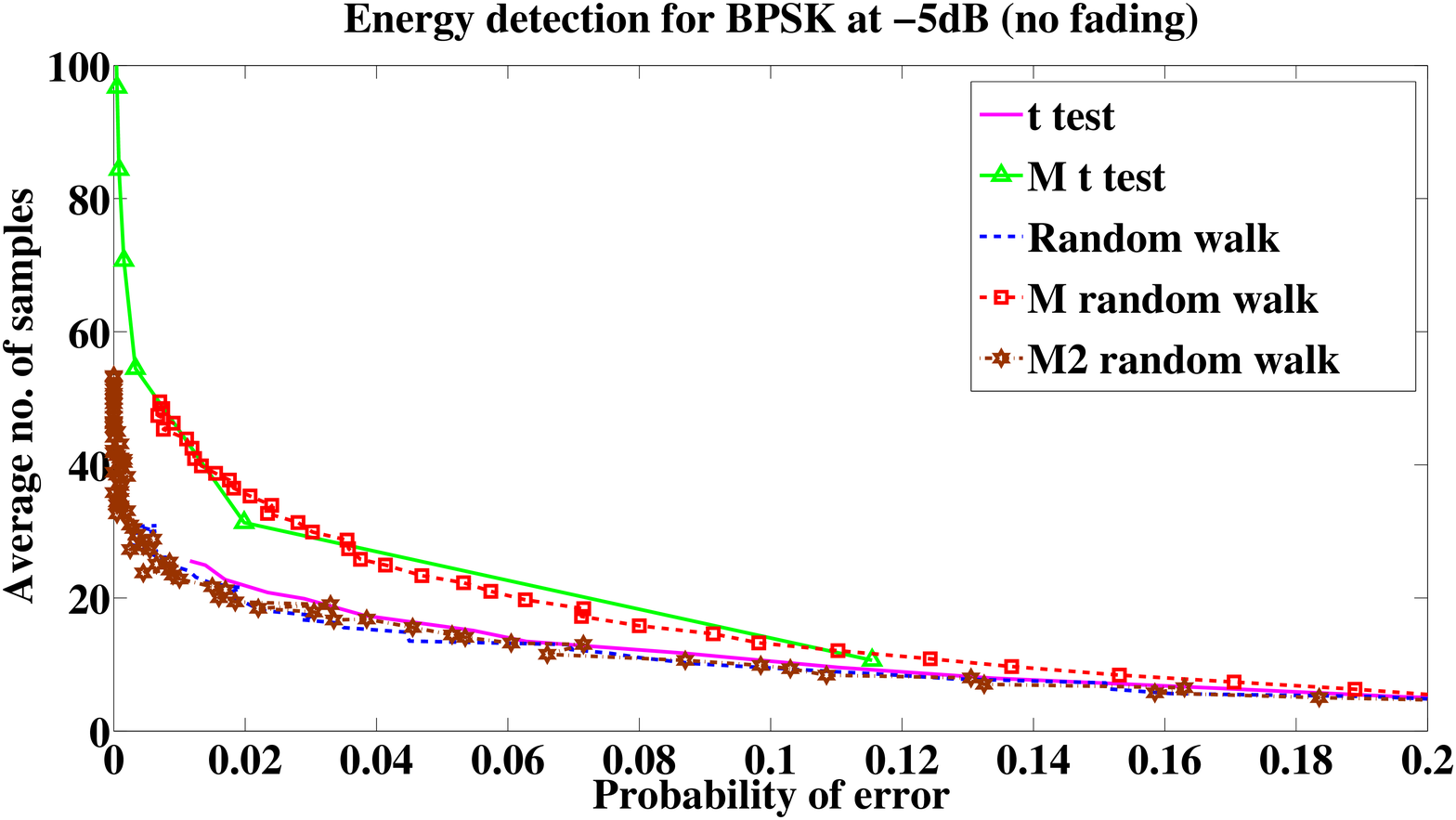}
		\end{subfigure}
        \begin{subfigure}[b]{0.5\textwidth}
                \centering
                \includegraphics[width=\textwidth,height=5cm]{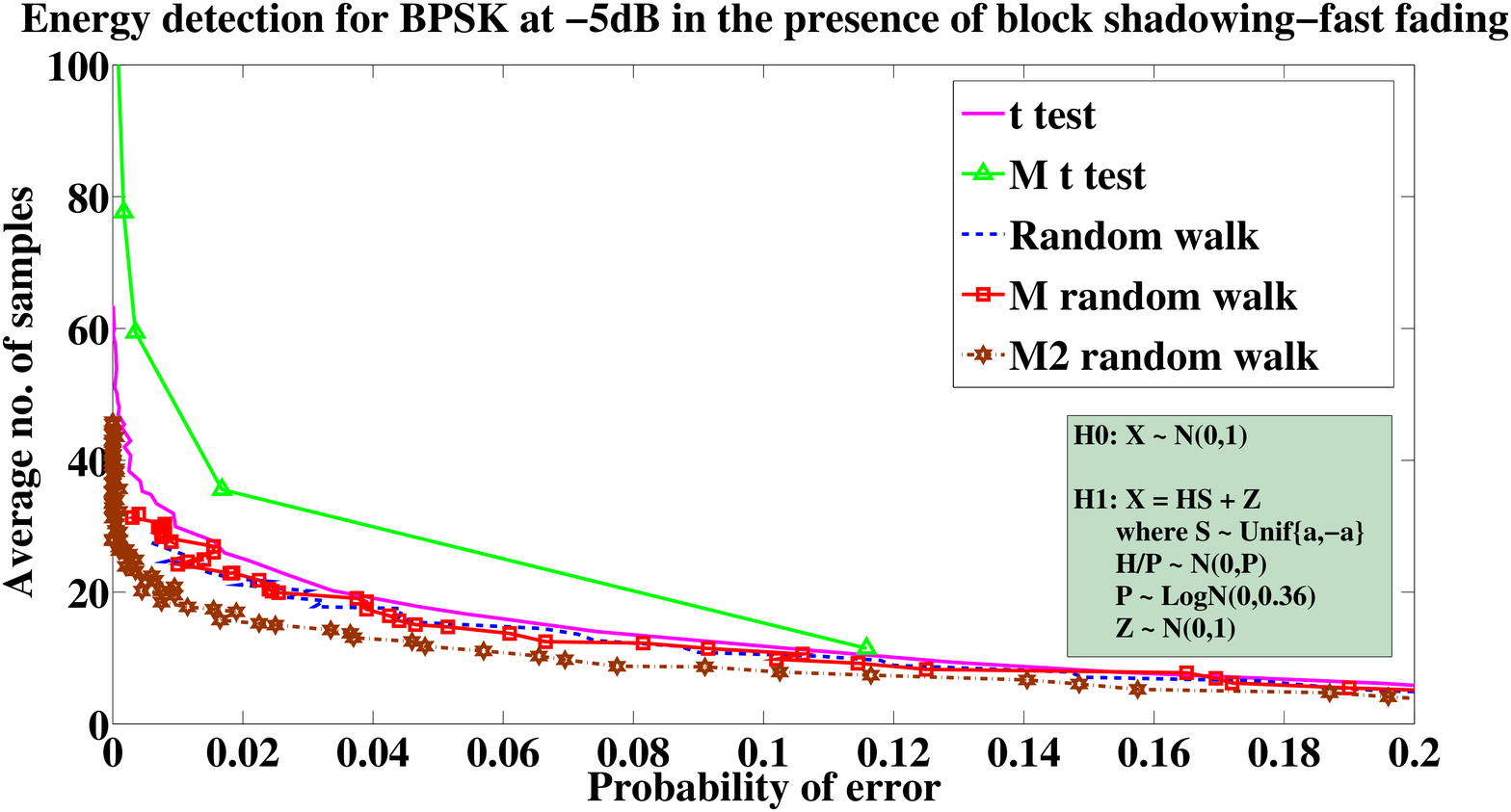}
                \end{subfigure}
\caption{\small Energy detection in the presence of Gaussian noise. Top: Without fading. Bottom: Under block Log $\mathcal{N}$ shadowing - Rayleigh fast fading.}\label{fig:energy_gaussian}
\end{figure}

\begin{figure}[ht]
        \centering
                \includegraphics[width=0.45\textwidth,height=5cm]{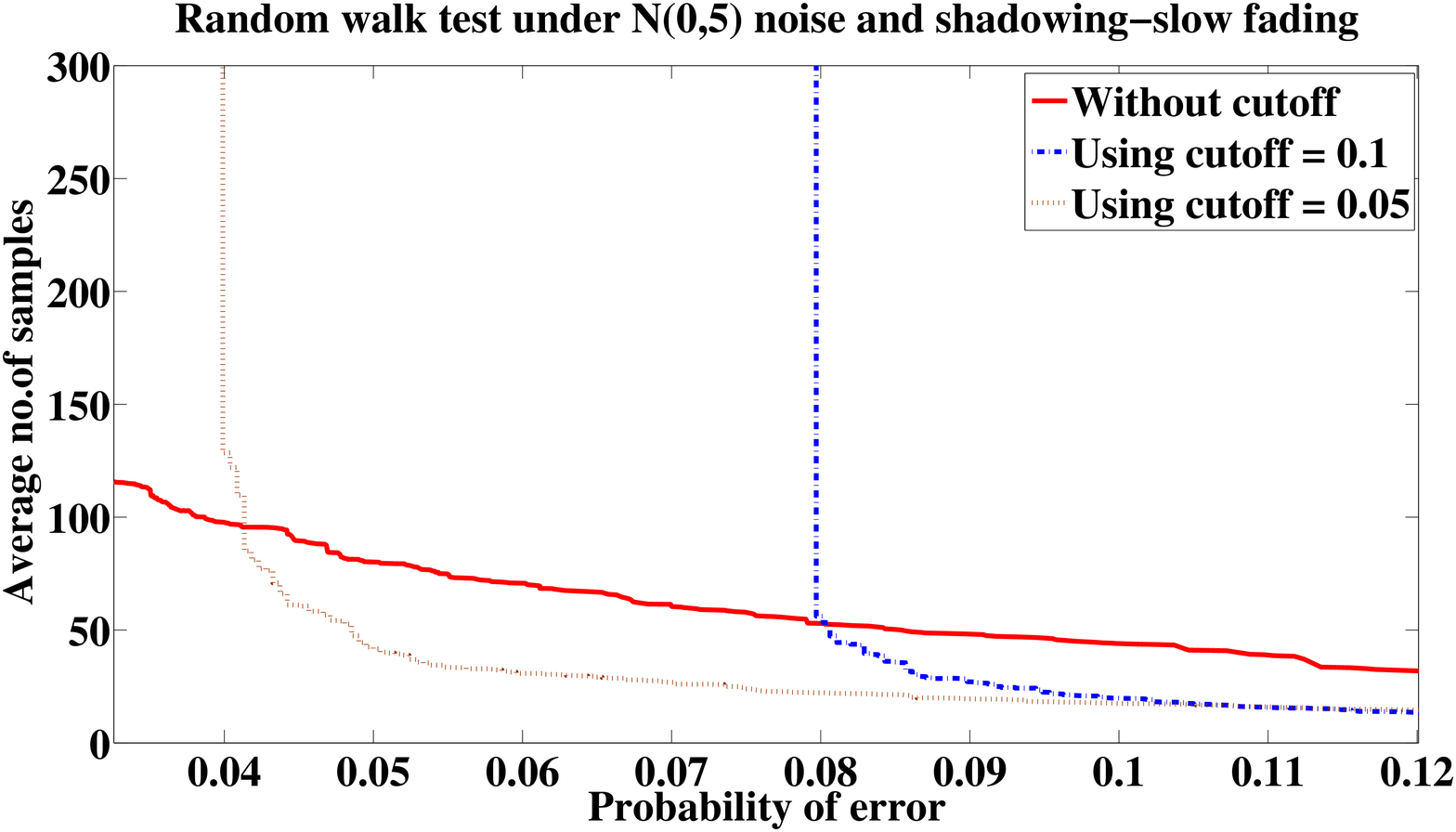}
\caption{\small Comparison of random walk with delta-random walk}\label{fig:delta_randomwalk}
\end{figure}

Figures~\ref{fig:compare_randomwalk}-~\ref{fig:energy_gaussian} show the simulations for various algorithms with different combinations of fast/slow fading, S$\alpha$S EMI and outliers. We draw the following conclusions.
\begin{itemize}
\item From Figures~\ref{fig:only_gaussian}-~\ref{fig:energy_gaussian} we see that the random walk test always performs better than the $t$ test and the rank test.
\item From Figures~\ref{fig:only_gaussian},~\ref{fig:gaussian_alpha},~\ref{fig:gaussian_alpha_outlier} comparing the top part of each figure (for fast fading) with the bottom part (for slow fading), for each algorithm, slow fading performs much worse. The effect of heavy-tailed EMI is somewhat like that of fast fading.
\item From Figure~\ref{fig:compare_randomwalk} we see that for random walk, slow fading has the most devastating effect on performance. This can be seen for other algorithms also from other figures. Next major damage is done by outliers. We see that heavy-tailed EMI also degrades the performance significantly.
\item From Figures~\ref{fig:only_gaussian},~\ref{fig:energy_gaussian} we observe that when there is only Gaussian noise and fast/slow fading $M$-random walk does not improve the performance over random walk. This is expected because the operation of $\psi$ is used only to improve the performance with respect to outliers and heavy-tailed EMI. We will see in the next section that degradation via (slow) fading is mainly due to the channel gain $H$ being low very often. Also see comments below.
\item That $M$-random walk and $M^2$-random walk are very effective in mitigating the effects of heavy-tailed EMI and outliers can be seen from Figures~\ref{fig:gaussian_alpha},~\ref{fig:gaussian_alpha_outlier}. From these we can conclude that outliers can cause major damage (for random walk, $t$ test) but are effectively handled by $M$-random walk. The rank test is not affected so much. In case of energy detection with S$\alpha$S EMI and fast fading, the only algorithm (among the algorithms considered) that works at all is $M^2$-random walk. Other algorithms do not provide probability of error $\le 0.3$.
\item Performance of $M^2$-random walk test with EMI is presented in Figure~\ref{fig:distri_SaS_fading} along with that of the distributed algorithm. From Figure~\ref{fig:distri_SaS_fading} we also see that unlike in Figure~\ref{fig:compare_randomwalk}, the outliers are helping the performance in the energy detection case. This is because we consider outliers only when there is signal ($\mathcal{H}_1$) and not under $\mathcal{H}_0$ unlike in Figure~\ref{fig:compare_randomwalk} where $\mathcal{H}_1$ and $\mathcal{H}_0$ both have signal.
\item As mentioned above, slow fading causes maximum degradation. This is because, for Rayleigh fading, the channel gain $H$ is low with a large probability. In that case, not making a decision when $H$ is very small is the sensible thing to do. Thus our algorithm $\delta$-random walk actually improves the performance significantly in this case (see Figure~\ref{fig:delta_randomwalk}).
\end{itemize}

\par Based on the above simulation results, we have decided to use the $M$-random walk at the FC and the $M^2$-random walk test at the local nodes. However, this happened because we took $\alpha$ in S$\alpha$S EMI as $1.8$. To allow for any $\alpha >0$ at the FC, we need to use $M^2$-random walk at the FC as well. $M^2$-random walk can be made to work close to $M$-random walk if we take $K$ in $\psi_1$ large. 
\par In the next section we will theoretically study these algorithms. Asymptotic analysis of the random walk test is provided in \cite{febi}. In the next section we briefly present that and also include the effects of heavy tailed noise and fading which was not discussed in \cite{febi}. This will explain why $M$-random walk and $M^2$-random walk perform better under heavy-tailed EMI and outliers and using truncation on $H$ improves performance in the presence of slow fading.

\section{Single Node: Analysis for random walk}
\label{section:single node analysis}
% Asymptotic analysis of the random walk test is provided in \cite{febi}. In this section we briefly present that and also include the effects of heavy tailed noise which was not discussed in \cite{febi}. This will explain why $M$-random walk and $M^2$-random walk perform better under heavy-tailed EMI and outliers. Then we will include the effect of fast and slow fading and show that $\delta$-random walk can improve the performance.
\par First we consider the scenario of mean detection. Here, the noise can have heavy tail due to Gaussian and symmetric $\alpha$-stable (or other heavy tailed) distribution. Furthermore, the fading distribution can also be heavy tailed. In the following we first provide the different classes of heavy-tailed distributions used in the analysis that follows.

\par The family of $\alpha$-stable laws is denoted by $S_\alpha(\sigma,\beta,\mu)$ \cite{SalphaS_survey} with $0 < \alpha \le 2$ its index, $\sigma > 0$ its scale parameter, $-1 \le \beta \le 1$ its skewness and $-\infty < \mu < +\infty$ its location. When $\alpha=2$ then it becomes $\mathcal{N}(\mu,2\sigma^2)$. All $\alpha$-stable laws have continuous, positive, uni-modal probability density function. A random variable $X$ with $\alpha$-distribution, $0 < \alpha < 2$ satisfies $P[X>x] \sim x^{-\alpha} \sim P[X<-x]$ and $\mathbb{E}[|X|^p]<\infty$ for $0<p<\alpha$ and $\mathbb{E}[|X|^p]=\infty$ for $p \ge \alpha$.

\par We also allow for the possibility of Gaussian mixture for EMI, which is light-tailed. Also, for $M$-random walk and $M-t$ test, due to bounded Huber $\psi$ function, all distributions become light-tailed.

\par We will use the following notation. For CDF $F$, $\overline{F}(x)=1-F(x)$, $F^{*2}$ is convolution of $F$ with itself and $\overline{F}^{*2}(x)=1-F^{*2}(x)$.
\par \emph{Definition }\cite{sigman}: $F$ is \emph{light-tailed} if $\int_{-\infty}^{\infty} e^{\alpha x}dF(x) < \infty$ for all $\alpha$ with $0 \le |\alpha| < \alpha_1$ for an $\alpha_1 \le \infty$; otherwise it is \emph{heavy-tailed}. $F$ is \emph{long-tailed} ($F \in \mathbb{L}$) if $\lim_{x \rightarrow \infty} \frac{\overline{F}(x+y)}{\overline{F}(x)}=1$ for all finite $y$. $F$ is \emph{sub-exponential} ($F \in \mathbb{S}$) if $\lim_{t \rightarrow \infty} \frac{\overline{B}^{*2}(x)}{\overline{B}(x)} = 2$ where $B$ is the distribution of $max\{0, X\}$ while $X$ has the distribution of $F$. $F$ is \emph{regularly varying} of index $-\alpha$, $\alpha \ge 0$, (denoted by $F \in R(-\alpha)$), if $\overline{F}(x) = l(x)x^{-\alpha}$, where $l$ is a slowly varying function, i.e., for all $\lambda > 0, \frac{l(\lambda x)}{l(x)} \rightarrow 1$ as $x \rightarrow \infty$. $F \in \mathbb{S}^*$ if $\lim_{t \rightarrow \infty} \int_0^t \frac{\overline{F}(t-x)}{\overline{F}(t)}\overline{F}(x)dx = 2 \int_o^\infty \overline{F}(x)dx$.

\par A long-tailed distribution is heavy-tailed. Also, $\mathbb{S}^* \subset \mathbb{S} \subset \mathbb{L}$ and $R(-\alpha) \subset \mathbb{S}$. If $F \in R(-\alpha)$ and it also has a finite mean, then it is in $\mathbb{S}^*$. Gaussian, exponential, Rayleigh and Laplace distributions are light-tailed while Pareto, log normal and Weibull distributions are sub-exponential. For $\alpha < 2, S_\alpha(\sigma,\beta,\mu)$ belongs to $R(-\alpha)$.  If $F \in R(-\alpha)$ then $\mathbb{E}[X^\beta] < \infty$ for $\beta < \alpha$ and $\mathbb{E}[X^\beta]=\infty$ for $\beta \ge \alpha$.

\par When $S_k$ takes values in a finite set and $H_k$ is light-tailed then $H_kS_k$ is light-tailed; if $H_k$ is heavy-tailed then $H_kS_k$ is heavy-tailed, if $H_k \in R(-\alpha)$ then $H_kS_k \in R(-\alpha)$. If independent random variables $X$ and $Y$ are light-tailed then $X+Y$ is light-tailed. If  any of $X$ and $Y$ is heavy-tailed so is $X+Y$. If $F \in \mathbb{S}$, $\bar{G}(x) = O(\bar{F}(x))$, then $F*G \in \mathbb{S}$. If $X$, $Y$ are long-tailed then $X+Y$ is long-tailed. If $X \in R(-\alpha_1)$, $Y \in R(-\alpha_2)$ then $(X+Y) \in R(-min\{\alpha_1, \alpha_2\})$. If $X \in \mathbb{L}$, then $X^2 \in \mathbb{L}$. If $X \in R(-\alpha)$, then $X^2 \in R(-\alpha/2)$.

\par The above results provide us the tail behaviour of $N_k + H_kS_k$, $N_k + H_kb_1$ and $N_k - H_kb_0$ in terms of tail behaviour of $N_k$ and $H_k$ where $b_0$ and $b_1$ are positive constants. We also see the effect of taking energy samples.

\par Consider the random walk statistics (\hspace{-.18cm}~\ref{equation:random}) or the robustified random walk (\hspace{-.12cm}~\ref{equation:M random}) with Huber function $\psi$.

We write it as $T_{n} = \sum_{k=1}^{n}Y_k$ where $Y_k = (X_{k}-\frac{\mu_{0} + \mu_{1}}{2})$ or $Y_k = \psi(X_{k}-\frac{(\mu_{0} + \mu_{1})}{2})$. We choose $\psi$ such that $\theta_0 \triangleq \mathbb{E}_0[Y_1] <0$ and $\theta_1 \triangleq \mathbb{E}_1[Y_1] >0$. Implications for $M^2$-random walk directly follow.

\par The sequential test for the random walk statistics stops at $N=inf\{n: Y_n \notin (-t_0, t_1)\}$ where $t_0, t_1 > 0$. We will discuss picking $t_0$ and $t_1$ later on. Once $t_0, t_1$ are fixed, the actual performance of the test does depend on the distribution of $Y_1$ and we study that now. Define, for $t>0$,
\begin{align*}
N_1(t) = inf\{n: T_n > t\}, N_0(-t) = inf\{n: T_n < -t\}.
\end{align*}
\par We consider $\mathbb{E}_0[N]$. The results will similarly hold for $\mathbb{E}_1[N]$. Let $M=\sup_{n \ge 0}T_n$.

Under $\mathcal{H}_0$, $\mathbb{E}_0[Y_k] = \theta_0 <0$. Thus $N_0(-t) < \infty$ a.s. for all $t>0$ and $\{N_1(t) = +\infty\} = \{M < t\}$ when $M<\infty$ a.s. Consider $N(t) = min\{N_0(t),N_1(t)\}$. Thus,
\begin{eqnarray*}
\lim_{t \rightarrow \infty} \mathbb{P}_0 [N(-t) = N_0(-t)] = 1, \text{ and}\\
\lim_{t \to \infty}\frac{N(-t)}{t}  = \lim_{t \to \infty}\frac{N_0(-t)}{t}, \text{ a.s.}
\end{eqnarray*}
Since we want to design algorithms with small probabilities of error, we will work with $t$ where $P[N(-t) = N_0(-t)]$ is large. Thus, we consider $N_0(-t_0)$. From random walk theory \cite{gut}, the following results hold. We have $\lim_{t \to \infty}\frac{N_0(-t_0)}{t_0} = \frac{-1}{\theta_0}$ a.s. and in $L_1$ even when $\theta_0 = -\infty$ (then the limit is $0$). For $r \ge 1$, if $\mathbb{E}[({Y_1}^-)^r] < \infty$ then $\mathbb{E}[(N_0(t_0)^r] < \infty$ and if $Y_1$ has finite moment generating function in a neighbourhood of $0$ then $N_0(t)$ also has. Here and in the following $Y_{1}^- = min\{0,Y_1\}$ and $Y_{1}^+ = max\{0,Y_1\}$. Also $F$ denotes the distribution of $Y_1$.
\par For $1<r<2$, if $\mathbb{E}[({Y_1}^-)^r] < \infty$ then $\mathbb{E}_0[N(t_0)] = \frac{t_0}{\theta_0} + o(t^{2-r})$. If $\mathbb{E}[({Y_1}^-)^2] < \infty$,
\begin{equation}
\frac{t_0}{\theta_0} \le \mathbb{E}[N_0(-t_0)] \le \frac{t_0}{\theta_0} + \frac{\mathbb{E}[({Y_1}^-)^2]}{2{\theta_0}^2} + o(1).
\label{equation:t0theta0}
\end{equation}
Similar results hold for $\mathbb{E}_1[N_1(t_1)]$ with conditions on $\mathbb{E}[({Y_1}^+)^r]$.
\par From above results we see that the tail behaviour of $F$ may not have much impact on $\mathbb{E}[N]$. For somewhat large $t_i$, $\mathbb{E}_i[N(t_i)]$ is close to $t_i/\theta_i, i=0,1$ under very weak conditions.

Next we consider $P_{FA}$. We have
\begin{equation}
P_{FA} = P_0[Y_N \ge t_1] \le P_0[M \ge t_1].
\label{equation:pfa}
\end{equation}

From \cite{asmussen}, if $\mathbb{E}[e^{\alpha Y_1}]<\infty$ for all $0 < \alpha < \alpha^* \le \infty$ and $\mathbb{E}[e^{\alpha Y_1}]=\infty$ for all $\alpha \ge \alpha^*$ then there exists a $\Gamma>0$ such that $\mathbb{E}[e^{\Gamma Y_1}] = 1$ and then
\begin{equation}
P_0[M\ge t_1] \le e^{-\Gamma t_1} \text{ for all } t_1 \ge 0.
\label{equation:p0t1}
\end{equation}
Also, if $Y_1$ is long tailed then
\begin{equation}
P_0[M\ge t_1] \sim \frac{1}{\theta_0}\bar{F}_I(t_1) \text{ as } t_1 \rightarrow \infty,
\label{equation:p0t1_long}
\end{equation}
where $\bar{F}_I(x) = {\int_{x}}^{\infty}(1-F(y))dy$ and $f(x) \sim g(x)$ denotes $\lim_{x \rightarrow \infty} \frac{f(x)}{g(x)} = constant$. Thus if $Y_1 \in \mathbb{R}(-\alpha)$ then $M \in \mathbb{R}(-\alpha+1)$ for $\alpha >1$ and if $F \in \mathbb{S}$, then $M \in \mathbb{S}$. For $M$-random walk, only (\hspace{-0.12cm}~\ref{equation:p0t1}) is relevant. From (\hspace{-0.12cm}~\ref{equation:pfa}), (\hspace{-0.12cm}~\ref{equation:p0t1}) and (\hspace{-0.12cm}~\ref{equation:p0t1_long}), we get an upper bound on $P_{FA}$ for light-tailed as well as long tailed distributions of $Y_1$. Because of our focus on $M$-random walk and $M^2$-random walk, light tailed case is of particular interest. If $P_{FA} \le \alpha$ is desired then from (\hspace{-0.12cm}~\ref{equation:p0t1}) we can get the threshold $t_1$ needed. However, $\Gamma$ depends on the distribution of $Y_1$. But approximations for $\Gamma$ are also available. For example, from \cite{asm_alb}, Chapter $IV$, $\Gamma < \frac{\mathbb{E}[Y_1]^2}{\mathbb{E}[Y_1^-]\mathbb{E}[Y_1^2]}$. This is a good approximation for $\mathbb{E}[Y_1]$ close to $0$, i.e.,  $\Gamma$ can be replaced with this upper bound. This bound depends only on the first two moments of $Y_1$ and, $\mathbb{E}[Y_1^-]$. Similarly we can use the $P_{MD} \le \beta$ to get $t_0$. These then provide $\mathbb{E}_0[N]$ and $\mathbb{E}_1[N]$.

Perhaps a more precise approximation of $P_{FA}$ can be obtained by observing that $P_{FA} = P_0[sup_{0\le k \le N_0(-t_0)}T_k > t_1]$. Since $N_0(-t_0)$ is a stopping time for the random walk $S_k$, if distribution $F$ of $Y_1 \in S^*$ then \cite{foss},
\begin{equation*}
\frac{P[sup_{0\le k \le N_0(-t_0)}S_k \ge x]}{1-F(x)} \rightarrow \mathbb{E}_0[N_0(-t_0)] \text{ as } x \rightarrow \infty.
\end{equation*}
Thus, if $t_1$ is somewhat large we can write
\begin{equation}
P_{FA} \sim (1-F(t_1))\mathbb{E}_0[N_0(-t_0)]
\label{equation:4}
\end{equation}
and use approximations and bounds on $\mathbb{E}_0[N_0(-t_0)]$ provided in (\hspace{-0.12cm}~\ref{equation:t0theta0}) and above it. Thus, $P_{FA}$ decays with $t_1$ at the same rate as the positive tail of $F$ as long as $F$ is in $S^*$. This provides a stronger result than (\hspace{-0.12cm}~\ref{equation:p0t1_long}): if $Y_1 \in R(-\alpha)$ then $P_{FA}(t_1) \sim t_1^{-\alpha}$. 

\par Similarly $P_{MD}$ depends on the negative tail of $F$. 
\par We use the above results to explicitly get the approximations for $\mathbb{E}_0[N]$ and $\mathbb{E}_1[N]$ for given $P_{FA} \le \alpha$ and $P_{MD} \le \beta$. For the light-tailed case, from (\hspace{-0.12cm}~\ref{equation:p0t1}) we get $t_1$ such that $e^{-\Gamma_0t_1} = \alpha$. Similarly we get $t_0$ such that $e^{-\Gamma_1t_0} = \beta$ where $\Gamma_0$ and $\Gamma_1$ are the $\Gamma$ coefficients in (\hspace{-0.12cm}~\ref{equation:p0t1}) under $P_0$ and $P_1$. For these $t_0$ and $t_1$, $\mathbb{E}_0[N] \approx \mathbb{E}_0[N_0(-t_0)] \sim \frac{t_0}{\theta_0} = \frac{1}{\theta_0\Gamma_1}|\log \beta|$ and $\mathbb{E}_1[N] \approx \mathbb{E}_1[N_1(t_1)] \approx \frac{1}{\theta_1\Gamma_0}|\log \alpha|$.

\par Now we consider the case where $1-F_0(t_1) \sim t_1^{-\alpha_1}$ and $1-F_1(t_0) \sim t_0^{-\alpha_1}$. Then $\alpha = P_{FA} \sim (1-F_0(t_1))\mathbb{E}_0[N(-t_0)] \sim t_1^{-\alpha_1} \frac{t_0}{\theta_0}$, $\beta = P_{MD} \sim t_0^{-\alpha_1} \frac{t_1}{\theta_1}$ and hence
\begin{equation}
\mathbb{E}_0[N] \sim {\theta_0}^{(\frac{1}{1-{\alpha_1}^2}-1)}(\alpha {\theta_1}^{\alpha_1} \beta^{\alpha_1})^{\frac{1}{1-{\alpha_1}^2}},
\label{equation:expected_0}
\end{equation}
\begin{equation}
\mathbb{E}_1[N] \sim {\theta_1}^{(\frac{1}{1-{\alpha_1}^2}-1)}(\beta {\theta_0}^{\alpha_1} \alpha^{\alpha_1})^{\frac{1}{1-{\alpha_1}^2}}.
\label{equation:expected_1}
\end{equation}
\par This shows that the performance of the random walk algorithm depends quite strongly on the tail behaviour of $F$ and with heavy tails the performance can really deteriorate.

\par Now we briefly comment of the performance of the (robust) random walk for mean detection: under $\mathcal{H}_0$, $Y_k = N_k - b_0H_k$ and under $\mathcal{H}_1$, $Y_k = N_k + b_1H_k$. We take $\mathbb{E}[N_k]=0$.

\par Initially assume that there is no fading. i.e., $H_k \equiv 1$. It is then a mean detection with $\mu_0 = -b_0$ and $\mu_1 = b_1$. Now the above analysis directly provides the effect of light and heavy-tailed $N_k$. Also, we see that by applying Huber function $\psi$ we can substantially gain in case of heavy-tailed $N_k$. For light-tailed case if we pick $K$ small, then it can make $\mu_0$ and $\mu_1$ smaller and hence one may see worse performance.

\par Next we consider the case of slow fading: $H_k \equiv H$. Now, it is realistic to assume that $H$ has been estimated and the receiver knows it (coherent detection case). Then we can consider observations $\overline{Y}_k = \frac{Y_k}{H} = \frac{N_k}{H} + b_1$ (or $\frac{N_k}{H} - b_0$). Since $N_k$ is zero mean, independent of $H$, $N_k/H$ stays zero mean. Also given $H=h$, $N_k/h$ will be heavy/light-tailed if $N_k$ is. Thus, it becomes the case considered in the previous paragraph. Denoting by $P_{FA}(t_0,t_1,h)$, $P_{MD}(t_0,t_1,h)$, $\mathbb{E}_0[N_0(-t_0,h)]$, $\mathbb{E}_1[N_1(t_1,h)]$ the corresponding quantities,  $\mathbb{E}_0[N_0(-t_0,h)] \approx \frac{t_0}{b_0}$, $\mathbb{E}_1[N_1(t_1,h)] \approx \frac{t_1}{b_1}$. For light-tailed case, $\mathbb{E}_0[N_0(H)] \approx \frac{1}{b_0\Gamma_1(H)}|\log \beta|$. If $N_k \sim \mathcal{N}(0,\sigma^2)$ then $\Gamma_1(H) = b_1H^2/\sigma^2$ and $\mathbb{E}_H[\mathbb{E}_0[N_0(H)]] \sim \frac{|\log \beta|\sigma^2}{b_0b_1}\mathbb{E}[\frac{1}{H^2}]$. For Rayleigh fading $\mathbb{E}[\frac{1}{H^2}] = \infty$. This is reflected in a significant performance degradation seen in the simulation results in Section~\ref{subsection:simu}.
\par If $P[N_k > t] \sim t^{-\alpha}$ then $P_0[\frac{N_k}{H} - b_0 > t_1] \sim ((t_1+b_0)h)^{-\alpha_1}$ and by (\hspace{-0.15cm}~\ref{equation:expected_0}) and (\hspace{-0.15cm}~\ref{equation:expected_1}) we get asymptotics for $\mathbb{E}_H[N_0(H)]$ and $\mathbb{E}_H[N_1(H)]$. We can further take expectation over $H$ to get the dependence on distribution of $H$. 
\par Above, we made the thresholds $t_0$ and $t_1$ dependent on $H$ and ensured that for each $h$, $P_{FA} \le \alpha$ and $P_{MD} \le \beta$. But this can often imply that $\mathbb{E}_H[N_1(H)]$ and/or $\mathbb{E}_H[N_0(H)] = \infty$. A weaker requirement is to choose $t_0$ and $t_1$ independently of $H$ such that $\mathbb{E}_H[P_{FA}(H)] \le \alpha$ and $\mathbb{E}_H[P_{MD}(H)] \le \beta$. It is possible that even now $\mathbb{E}_H[N_1(H)]$ and/or $\mathbb{E}_H[N_0(H)] = \infty$. In that case we can find positive constants $\delta_1, \delta, \alpha', \beta'$ such that $\delta_1 < \min\{\alpha,\beta\}$ and $P[|H|<\delta] \le \delta_1$ with $\mathbb{E}[P_{FA}(H)|H \ge \delta] \le \alpha'$, $\mathbb{E}[P_{MD}(H)|H \ge \delta] \le \beta'$ and
\begin{equation}
\begin{aligned}
\mathbb{E}[P_{FA}(H)] &= P[|H| \le \delta] + \mathbb{E}[P_{FA}(H)|H \ge \delta]P(H \ge \delta)\\
&\le \delta_1 + \alpha'(1-\delta_1) = \alpha
\label{equation:exp_delta}
\end{aligned}
\end{equation}
and $\delta + \beta'(1-\delta) = \beta$. Now we do not make a decision when $|H| \le \delta$. For this case we can ensure that $\mathbb{E}_H[N_i(H)] < \infty$ for $i=0,1$. At least for Gaussian $N_k$ and Rayleigh fading example above, $\mathbb{E}[\frac{1}{H^2}||H| \ge \delta] < \infty$. %Simulations in Section~\ref{subsection:simu} confirm that this technique provides a much better performance for slow fading case.

\par If we assume that we only know the sign of $H$ and not its magnitude (partial coherence -- knowing the phase only) then we define $\overline{Y}_k = sgn(H)Y_k = sgn(H)N_k + |H|b_1$ under $\mathcal{H}_1$ (or $sgn(H)N_k - |H|b_0$ under $\mathcal{H}_0$). From the distribution of $N_k$, we get the distribution of $sgn(h)N_k$ and obtain the asymptotics of our performance measures. In particular, if $N_k$ is zero mean, symmetric, $sgn(h)N_k$ has the same distribution as $N_k$. Also $\mathbb{E}_0[N_0(-t_0,h)] \approx \frac{t_0}{|h|b_0}$,  $\mathbb{E}_1[N_1(t_1,h)] \approx \frac{t_1}{|h|b_1}$ and if $P[N_k > t] \sim t^{-\alpha}$ then we can get from (\hspace{-0.15cm}~\ref{equation:expected_0}) and (\hspace{-0.15cm}~\ref{equation:expected_1}), $\mathbb{E}_1[N(H)]$ and $\mathbb{E}_0[N(H)]$.
\par From the above two paragraphs, we can see the advantage of knowing the magnitude $|H|$ at the receiver. Also, not knowing $|H|$ implies that we cannot decide when $|H| \ge \delta_1$ as needed in (\hspace{-0.12cm}~\ref{equation:exp_delta}). Analysis of $P_{MD}$ follows in the same way.
\par If the phase of $H$ is also not known, then random walk algorithm is not the right choice for this problem because it will perform quite badly.

\par Now we consider the fast fading case where $\{H_k$\} is i.i.d. This is a less likely scenario but we briefly discuss it because it leads to some new results. As above, if we have a noncoherent case (no sign or magnitude of $H$ available) then we should not use the random walk algorithm. The case of coherent detection (phase and magnitude both available) seems quite unlikely. Thus we consider partial coherence case where only the sign of $H$ is available. Taking $\overline{Y}_k = sgn(H_k)Y_k = sgn(H_k)N_k + |H_k|b_1$ under $\mathcal{H}_1$ and $\overline{Y}_k = sgn(H_k)N_k - |H_k|b_0$ under $\mathcal{H}_0$,we obtain the following conclusions:

\begin{itemize}
\item If $N_k$ has light positive and negative tails, but $H_k$ is heavy-tailed, $P_1$ has a positive heavy tail and light negative tail and vice versa for $P_0$. Thus, system performance is not affected by the heavy-tailed $H_k$. One can see some beneficial effects because $\mathbb{E}_i[N_i]$ will be somewhat shorter which is not captured by our analysis.
\item If $N_k$ has heavy positive and negative tails, but $H_k$ is light-tailed then $P_0$ and $P_1$ both have heavy positive and negative tails. Thus, $P_{FA}$ and $P_{MD}$ both suffer.
\item If $N_k$ and $H_k$ both are heavy-tailed then again $P_{FA}$ and $P_{MD}$ suffer.
\end{itemize}

\par Now we consider the system described in Section~\ref{section:model}. Under $\mathcal{H}_0$, $\tilde{X}_k = N_k$ and under $\mathcal{H}_1$, $\tilde{X}_k = H_kS_k + N_k$. As discussed, we use energy detection for this case by taking samples $X_k$ in (\hspace{-0.18cm}~\ref{equation:energy}). Then, from the results above, if $\{S_k\}$ is i.i.d. with values in a finite set and $\{H_k\}$ is i.i.d. (fast fading) depending on the tail behaviour of $H_k$ and $N_k$, we know the tail behaviour of energy samples $X_k$. Also, under various SNR conditions, we know that the energy detection problem can be considered the mean detection problem and the above results can be directly used. We do not need any information about $H_k$ itself; only the mean of $X_k$ under $\mathcal{H}_1$ and $\mathcal{H}_0$ may be required (at least for the low SNR case).
\par For slow fading case, $H_k \equiv h$, a constant in the sensing duration. Then, at low SNR, it is mean detection with $\mu_0 = M\sigma^2$ and $\mu_1 = M(\sigma^2 + h^2\mathbb{E}[S_k^2])$. Now, for given thresholds $-t_0$ and $t_1$, $\mathbb{E}[N_0(-t_0,h)] = \frac{t_0}{M\sigma^2}$ and $\mathbb{E}[N_1(t_1,h)] = \frac{t_1}{M(\sigma^2 + h^2\mathbb{E}[S_k^2])}$. Also, $P_{FA}(t_0,t_1,h)$ and $P_{MD}(t_0,t_1,h)$ can be approximated/bounded as above and the effect of heavy and light-tailed $N_k$ can be studied. Taking expectation over $H$ will provide the effects of tail of the distribution of $H$ as well.

\par If $H_k \equiv H$ (slow fading) and unknown, then let for $H=h$, $P_{FA}(h)$, $P_{MD}(h)$, $\mathbb{E}[N_0(-t_0, h)]$, $\mathbb{E}[N_1(t_1, h)]$ represent the corresponding probabilities of error and expected detection times. Then $\mathbb{E}[N(-t_0)] \approx t_0/h$. If $N_k \in \mathbb{S}^*$, then $P_{FA}(t_1,h) \sim (1-F_0(t_1 + ht_0))\frac{t_0}{h}$ where $F_0$ is the cdf of $N_k$. Also, $\mathbb{E}_H[P_{FA}(t_1,H)] \sim \int_0^\infty (1-F_0(t_1 + ht_0))\frac{t_0}{h} dP_H(h)$ where $P_H$ is the distribution of $H$. Similarly one can study the case of $H$ being light-tailed. The analysis for $P_{MD}$ is along the same lines. In this case particularly, since $\psi_0$ and $\psi_1$ are bounded, one expects that $M$-random walk and $M^2$-random walk will provide much better performance. 
\par This study explains the results observed in Section~\ref{subsection:simu}.

\section{Asymptotic Analysis}
\label{section:distributed analysis}
Based on the simulation results in Section~\ref{section:survey} and the theory in Section~\ref{section:single node analysis} we now consider the distributed algorithm where each local node and the FC use $M^2$-random walk. In addition, we also use $\delta$-truncation. We call this distributed algorithm, \emph{$M^2$-$M^2$-$\delta$-random walk}. Exact theoretical analysis of this algorithm is intractable. Therefore, in this section we provide an asymptotic analysis of the algorithm which provides the performance as the $P_{FA}$ and $P_{MD}$ tend to zero. This analysis provides good insight but does not provide a good approximation of the algorithm at practical parameter values. Thus in the next section we will also present an approximation analysis which provides a much better approximation to the performance at usual parameters of interest than the asymptotic results provided here. 
\par The observations at the local nodes and the fusion node after operation with the $\psi$ function are light-tailed, in fact bounded. Therefore, assumptions of Theorem $2$ and $3$ below, will be satisfied. Comparing Theorem $3$ with Theorem $4$ shows the advantage of using $\psi_0$ and $\psi_1$. The following analysis is not affected by $\delta$-truncation.

\par Let
\begin{align*}
\hat{X}_{1l} = \psi_0\Big(\psi_1(X_{1l})-\frac{\mu_{0l}+\mu_{1l}}{2}\Big),
\end{align*}
where $\mathbb{E}_1[\psi_1(X_{1l})] \ge \mu_{1l}, \mathbb{E}_0[\psi_1(X_{1l})] \le \mu_{0l}$ and, $\mu_{1l} > \mu_{0l}$ for $l=1,2,..., L$.

We choose $\psi_{0}$ and $\psi_{1}$ such that under $\mathcal{H}_0, \mathbb{E}[\hat{X}_{1l}] < 0,$ under $\mathcal{H}_1, \mathbb{E}[\hat{X}_{1l}] > 0$. Then, $T_{nl}=\sum_{k=1}^n \hat{X}_{kl}$ and $W_n = \sum_{k=1}^n \psi_0 \Big(\psi_1(Y_k) - \frac{\overline{\mu}_0 + \overline{\mu}_1}{2}\Big)$ where $0>\overline{\mu}_0$ and $0<\overline{\mu}_1$ are selected properly such that $\overline{\mu}_0 \ge -b_0L, \overline{\mu}_1 \le b_1L$. Let $\hat{Z}_k = \psi_0 \Big(\psi_1(Y_k) - \frac{\overline{\mu}_0 + \overline{\mu}_1}{2}\Big)$.
\par We use the following notation:
\begin{align*}
 & \Delta_i = \text{ mean drift of }W_k\text{ when all local nodes decide }\mathcal{H}_i,\\
 & D_{tot}^i = \sum_{l=1}^L \mathbb{E}_i\big[ \hat{X}_{kl}\big],\\
 & N = \inf\{k: F_k \ge \beta_1 \text{ or } F_k \le -\beta_0\},\\
 & \xi_i^* = \psi_0(\psi_1(Lb_1 + Z_i) - \frac{\overline{\mu}_0 + \overline{\mu}_1}{2}),\\
 & \xi_i^{**} = \psi_0(\psi_1(-Lb_0 + Z_i) - \frac{\overline{\mu}_0 + \overline{\mu}_1}{2}),\\
 & R_i^l = -\log \inf_{t \ge 0} \mathbb{E}_i\big[e^{(-1)^i t (\hat{X}_{il}-\frac{\theta_{0l}+\theta_{1l}}{2})}\big].\\
  & R_i = \min_l R_i^l.\\
  & \text{We choose } \psi_0 \text{ such that } \Delta_0 < 0 \text{ and } \Delta_1 > 0.
 \end{align*}

% & {N_l}^1 = inf\{k: T_{kl} \ge \gamma_{1l}\}\\
% & {N_l}^0 = inf\{k: T_{kl} \le -\gamma_{0l}\}\\
% & N_l = \min\{{N_l}^0,{N_l}^1\}\\
% & {N}^1 = inf\{k: F_{k} \ge \beta_{1l}\}\\
% & {N}^0 = inf\{k: F_{k} \le -\beta_{0l}\}\\

% & \zeta_i^* = \text{mean drift of random walk at FC when all local nodes } \\
% & \hspace{0.8cm}\text{decide} H_i\\
% & \tau_l(-\gamma_{0l})= \sup\{k:  T_{kl} \ge -\gamma_{0l}\}\\
% & \tau = \sup\{\tau_l(-\gamma_{0l}), l=1,..., L\}\\
% & \nu (a) = \text{ first time } W_k \text{ crosses } a \text{ when all nodes make correct}\\
% & \hspace{1cm}\text{ decisions}.
%Under $H_0$,
%\begin{align*}
%& \frac{N}{|log c|} \le \frac{N_0}{|log c|} \le \frac{\tau(c)}{|log c|} + \frac{\nu(-|log c|-F_{\tau(c)+1}}{|log c|}\\
%& \le \frac{\tau(c)}{|log c|} + \frac{\nu(-|log c|}{|log c|}+ \frac{\nu(-F_{\tau(c)+1}}{|log c|}\\
%\end{align*}
%
%
%\begin{align*}
%& \lim_{t \to \infty}\frac{\nu(-t)}{t} = \frac{1}{\Delta_0} a.s.\\
%& \lim_{t \to \infty}\frac{\nu(-F_{\tau(c)}+1}{|log c|} =  \lim_{\tau \to 0}\frac{\nu(-F_{\tau(c)}+1}{F_{\tau(c)}+1}.\frac{F_{\tau(c)}+1}{|log c|}\\
%& = \frac{1}{\Delta_0}.\mathbb{E}_0[Y_1]L
%\end{align*}

\par $\mathbf{Theorem \hspace{0.2cm} 1:}$ For any finite thresholds $\gamma_{il}$, $\beta_i$, $P_{i}[N < \infty] = 1$.\\

$\mathbf{Proof: }$ Please see the appendix. \hfill $\square$\\
\par For Theorems $2$-$4$, we will use the following thresholds:
\begin{align*}
 & -\beta_0 = -|\log c|, \beta_1 = |\log c|,\\
 & \gamma_{0l}=-\gamma_l|\log c|, \gamma_{1l}=\rho_l|\log c|, \text{ where,}\\
 & \gamma_l = \frac{\mathbb{E}_0\big[\hat{X}_{kl}\big]}{D_{tot}^0}, \rho_l = \frac{\mathbb{E}_1\big[\hat{X}_{kl}\big]}{D_{tot}^1}.
\end{align*}
$\mathbf{Theorem \hspace{0.2cm} 2:}$ Let $\mathbb{E}_i[|\hat{X}_{1l}|^{\alpha+1}] < \infty$ for $l=1,2,...,L$ and $\mathbb{E}_i[|\hat{Z}_1|^{\alpha+1} < \infty]$ for some $\alpha > 1$. Then under $\mathcal{H}_i$,
\begin{equation*}
\limsup_{c \rightarrow 0} \frac{N}{|\log c|} \le \frac{1}{D_{tot}^i} + M_i \text{ a.s.}
\end{equation*}
and in $L_1$ where $M_i = \frac{c_i}{\Delta_i}, c_0 = -\big[1+\frac{\mathbb{E}_0|\xi_1^*|}{D_{tot}^0}\big], c_1 = \big[1+\frac{\mathbb{E}_1|\xi_1^*|}{D_{tot}^1}\big]$.  \\

$\mathbf{Proof: }$ Please see the appendix. \hfill $\square$\\

%\par The following result provides the rate of convergence of $P_{FA}$ and $P_{MD}$ to zero as the local and FC thresholds tend to $\infty$ (and $-\infty$) for the light-tailed $F$. Let $g$ be the moment generating function of $|\zeta_1^*|$ and $\Lambda(\alpha) = \sup_\lambda(\alpha \lambda - \log  g(\lambda)), \alpha^+ = \text{ ess }\sup |\zeta_1^*|$. Also, let
%\[ s(\eta) = \left \{ 
%\begin{array}{l l}
%\frac{\eta}{\alpha ^+} & \quad ,\text{ if } \eta \ge \Lambda(\alpha^+), \\
%\frac{\eta}{\Lambda^{-1}(\eta) } & \quad ,\text{ if } \eta \in (0,\Lambda(\alpha^+)).
%\end{array}\right. \]\\ 
We make the following assumptions for the next theorem.
\begin{itemize}
\item $\mathbb{E}_i[e^{\alpha_l\hat{X}_{1l}}] < \infty$ for $|\alpha_l| < \alpha_l^* \le \infty$ and $\mathbb{E}[e^{\alpha_l^*\hat{X}_{1l}}] = \infty$ for some $\alpha_l^* \le \infty$, for $i=0,1$. This implies that there exist $\Gamma_{il} > 0, i=0,1, l=1,...,L,$ such that  $\mathbb{E}[e^{\Gamma_{il}\hat{X}_{il}}]=1$ (\cite{asm_alb}).
\item There exists $\alpha_0 > 0$ such that $\phi_{\xi^*}(\alpha_0) \triangleq \mathbb{E}_0[e^{\alpha_0{\xi^*}}] < \infty$, and a $\beta_0 > 0$ such that $\phi_{\xi^{**}}(\beta_0) \triangleq \mathbb{E}_1[e^{\beta_0{\xi^{**}}}] < \infty$.
\item For $k_2 \triangleq \sum_l \gamma_l \gamma_l'$ where $\gamma_l'$ is the smallest positive constant with $\mathbb{E}[e^{-\gamma_l'\hat{X}_{1l}}] = e^{-\eta}$ for all $l=1,..., L$ and $\eta$ is some positive constant less than $R_0$, $k_2 < \alpha_0$. Also let $\log \phi_{\xi^*}(\alpha_0) \le \eta$. Similarly we define conditions for $\mathcal{H}_1$. 
\item There exist constants $\Gamma_0, \Gamma_1 > 0$ such that $\mathbb{E}_i[e^{\Gamma_i\hat{Z}_1}] = 1$, for $i=0,1$.
\item There is $\alpha_1 >0$ such that $\bar{\phi}_{il}(\alpha_1) \triangleq \mathbb{E}[e^{-\alpha_1\hat{X}_{1l}}] < \infty$ for all $l=1,..., L$ and $\eta + \log \bar{\phi}_{0l}(\alpha_1) < \alpha_1\mathbb{E}[-\hat{X}_{1l}] < 0$. Also, there is $\beta_1 >0$ such that $\bar{\phi}_{il}(\beta_1) \triangleq \mathbb{E}[e^{-\beta_1\hat{X}_{1l}}] < \infty$ for all $l=1,..., L$ and $\eta + \log \bar{\phi}_{0l}(\beta_1) < 0$.
\item $\mathbb{E}[\xi_1^*] \ge 0$, $\mathbb{E}[\xi_1^{**}] \le 0$.\\
\end{itemize}

\par $\mathbf{Theorem \hspace{0.2cm} 3:}$ Under the above assumptions, \begin{enumerate}\item[(a)] $\lim_{c \downarrow 0} \frac{P_{FA}}{c^{r'}} < \infty$ for any $r'$, with $0<r'<\min\{r\alpha_0-k_2, \Gamma_0(1-r), \Gamma_{0l}\gamma_l, l=1,..., L\}$ for some $0<r<1$.
\item[(b)]$\lim_{c \downarrow 0} \frac{P_{MD}}{c^{s'}} < \infty$ for any $s'$, with $0<s'<\min\{s\alpha_1-k_2', \Gamma_1(1-s), \Gamma_{1l}\gamma_l, l=1,..., L\}$ for some $0<s<1$.\\
\end{enumerate}
\par $\mathbf{Proof: }$ Please see the appendix. \hfill $\square$\\

\par We verify the above assumptions for the Gaussian distribution. Then we do not use $\psi_0$ or $\psi_1$. Thus, $X_{1l} \sim \mathcal{N}(\mu_{0l},\sigma_l^2)$ under $\mathcal{H}_0$ and $X_{1l} \sim \mathcal{N}(\mu_{1l},\sigma_l^2)$ under $\mathcal{H}_1$. Also, $\hat{Z}_1 \sim \mathcal{N}(\frac{\overline{\mu}_0-\overline{\mu}_1}{2}, \overline{\sigma}^2)$ under $\mathcal{H}_0$ and $\hat{Z}_1 \sim \mathcal{N}(\frac{\overline{\mu}_1-\overline{\mu}_0}{2}, \overline{\sigma}^2)$ under $\mathcal{H}_1$.
\par Now, $R_0^l = \frac{1}{2}\frac{\mu_{0l}^2}{\sigma_l^2}$. Assuming that the means and variances are the same at each node, i.e., $\mu_{il} = \mu_i$ and $\sigma_{il}^2 = \sigma^2$ for $i=0,1$, we get $R_0 = \frac{1}{2}\frac{\mu_0^2}{\sigma^2}$. Now $\log \phi_{\xi_1^*}(\alpha_0) = \mu_0\alpha_0 + \frac{1}{2}\sigma^2\alpha_0^2$. We need to check if there exists an $\eta < R_0$ such that $\log \phi_{\xi^*}(\alpha_0) \le \eta$. Thus, we need to find $\alpha_0$ such that $\log \phi_{\xi^*}(\alpha_0) < \frac{1}{2}\frac{\mu_0^2}{\sigma^2}$. This translates to finding $\alpha_0$ such that $\alpha_0 < \frac{-\mu_0-\sqrt{\mu_0^2+\mu_0}}{\sigma^2}$. Now, by definition, $k_2 \triangleq \sum_l \gamma_l \gamma_l'$ where $\gamma_l' = \min\{\gamma > 0: \mathbb{E}[e^{\gamma\hat{X}_{1l}}] = e^{-\eta}\}$. Thus $\gamma_l' = \frac{\mu_1 - \sqrt{\mu_1^2 - 2\sigma^2\eta}}{\sigma^2}$. From the definition of $\gamma_l$, we get $\gamma_l = \frac{1}{L}$ where $L$ is the number of local nodes. Thus, $k_2 = \frac{1}{\sigma^2}(\mu_1 - \sqrt{\mu_1^2 - 2\sigma^2\eta})$. We need to check if $k_2 < \frac{-\mu_0-\sqrt{\mu_0^2+\mu_0}}{\sigma^2}$ so that a choice of $\alpha_0$ and $k_2$ satisfying $k_2 < \alpha_0$ is possible. This is equivalent to checking if $\mu_1^2 < \mu_1^2 - 2\sigma^2\eta$, which holds true for any positive $\eta$. Thus, we can choose any $\eta$ such that $0<\eta<R_0$ and $\eta + \log \bar{\phi}_{0l}(\alpha_1) < \alpha_1\mathbb{E}[-\hat{X}_{1l}] < 0$. We also note that $\Gamma_0 = \Gamma_1 = \frac{\overline{\mu}_1 - \overline{\mu}_0}{\overline{\sigma}^2}$ are the positive constants satisfying $\mathbb{E}_i[e^{\Gamma_i\hat{Z}_1}] = 1$, for $i=0,1$.

\par The following result is for heavy-tailed case. This is provided to show that if we do not robustify the observations at the local nodes and/or FC, the penalty for heavy-tailed EMI/outliers can be high. This holds for single node case also as demonstrated in Section~\ref{section:single node analysis}. For the following theorem, we work with the random walk algorithm ($\hspace{-.12cm}~\ref{equation:random}$).\\

$\mathbf{Theorem \hspace{0.2cm} 4:}$ If there is an $r_1 > 1$ and $r_2 >0$ such that the distribution of $X_{1l} \in R(-r_1)$ for all $l=1,2,...,L$ and under $\mathcal{H}_0$ and $\mathcal{H}_1$ and the distribution of $Z_1 \in R(-r_2-1)$ then
\begin{equation*}
P_{FA} \le o(|\log c|^{-min\{r_1,r_2\}+\epsilon}),
\end{equation*}
\begin{equation*}
P_{MD} \le o(|\log c|^{-min\{r_1,r_2\}+\epsilon}).
\end{equation*}
for any $\epsilon > 0$.\\
$\mathbf{Proof: }$ Please see the appendix. \hfill $\square$
%\par Thus we see that heavy-tailed noise/EMI affects the performance of $P_{FA}$ and $P_{MD}$.

\section{Approximation Analysis}
\label{section:approx}
In this section we provide an approximation analysis of the algorithm.

\par In the following, we take, for convenience, $b_{1}=-b_{0} = b$, and $\mu_{1} = -\mu_{0} = \mu = I.b$, for some $I$ with $1\leq I\leq L$. Roughly speaking, this ensures that the FC makes decision $\mathcal{H}_{1}$ when $I$ more nodes decide $\mathcal{H}_{1}$ compared to the nodes deciding $\mathcal{H}_{0}$. Similarly for $\mathcal{H}_{0}$. 
\begin{equation*}N_l^{1}\triangleq \inf\{n:\mbox{ }T_{nl} \ge \gamma_{1l}\}, N_l^{0}\triangleq \inf\{n:\mbox{ }T_{nl} \le \gamma_{0l}\}, 
\end{equation*}
\begin{equation*}
N_l = \min\{N_l^1,N_l^0\}
\end{equation*}
Similarly, $N^{1}$, $N^{0}$ and $N$ represent the corresponding terms for the FC. 

\par From Theorems $3$ and $4$ we know that as $\gamma_{0l}, \gamma_{1l} \to \infty$ and $\beta_0, \beta_1 \to \infty$, $P_{FA}, P_{MD} \to 0$. One can similarly show that as $\gamma_{0l}, \gamma_{1l} \to \infty$, the local decisions made by each local node are correct with probability $1$.
\par We will use the following notation:\\
$\delta_{i,FC}^{j}\triangleq$ mean drift of the FC process $\{W_{k}\}$ under $\mathcal{H}_{i}$, when $j$ local nodes are transmitting.\\
$t_{j}\triangleq$ time at which the mean drift of $\{W_{k}\}$ changes from $\delta_{i,FC}^{j-1}$ to $\delta_{i,FC}^{j}$.\\
$\tilde{W}_{j}\triangleq \mathbb{E}[W_{t_{j}-1}].$
\par Under $\mathcal{H}_{i}$,\begin{equation*}\tilde{W}_{j} = \tilde{W}_{j-1} + \delta^{j-1}_{i, FC}(\mathbb{E}(t_{j}) - \mathbb{E}(t_{j-1})), \tilde{W}_{0} = 0.
\end{equation*}
Based on the fact that $P_{FA}$ and $P_{MD}$ of each local node $l \to 0$ as $\gamma_{0l}, \gamma_{1l} \to \infty$ for each $l$, we get\\

$\mathbf{Lemma\mbox{ }1}$.  $P_{i}(\mbox{decision of the local node at time }t_{k}\mbox{ is }\mathcal{H}_{i}$ and $t_{k}\text{ is the }k^{th}\text{ order statistics of }\{N_{1}^{i},..., N_{L}^{i}\}) \to 1$ as $\alpha_{l},\beta_l \to 0,\mbox{ }\forall \mbox{ }l$. \hfill$\blacksquare$\\ \\
$\mathbf{Lemma \mbox{  }2}$.  Under $\mathcal{H}_{0}$, when $\alpha_{l}$ and $\beta_{l}$ are small, 
\begin{equation*}\displaystyle N_{l}^{0} \sim \mathcal{N}(\frac{-|\gamma_{0l}|}{\delta_{0,l}},\frac{-|\gamma_{0l}|\rho_{0,l}^{2}}{\delta_{0,l}^{3}}),\end{equation*}
where $\delta_{0,l}\displaystyle\triangleq \mathbb{E}_{0}[\hat{X}_{k,l}]$, and $\rho_{0,l}^{2}\displaystyle\triangleq$variance of $[\hat{X}_{k,l}]$ under $\mathcal{H}_{0}$.\\
\textbf{Proof}: See Theorem $5.1$, Chapter $3$ in \cite{gut}.\hfill $\blacksquare$\\
\par A similar result holds for $\mathcal{H}_{1}$ as well.
\par Based on the above lemmas, in the following we provide an approximation for $\mathbb{E}_{i}[N], i=0,1$. 

%\par When $\alpha_{l}, \beta_l$ and $\alpha, \beta$ are small, probabilities of error are small, as proved in the above lemmas. Hence in such a scenario, for approximation, we assume that local nodes are making correct decisions. 
\par Let,\\
\begin{equation*}
l_{0}^{*}\triangleq\min \{j: \delta^{j}_{0, FC} < 0 \text{ and }\frac{\gamma_{0l} - \tilde{W_{j}}}{\delta^{j}_{0, FC}}< \mathbb{E}(t_{j+1}) - \mathbb{E}(t_{j})\}.  
\end{equation*}
Then we can have the approximation
\begin{equation}
\mathbb{E}_0[N] \approx \mathbb{E}(t_{l_{0}^{*}}) + \frac{\gamma_{0l} - \tilde{W}_{l_{0}^{*}}}{\delta^{l_{0}^{*}}_{0, FC}}.
\label{eqn:approx_e0}
\end{equation}
The first term in  approximation (\hspace{-0.15cm}~\ref{eqn:approx_e0}) corresponds to the mean time till the mean drift of $\{W_{k}\}$ becomes  negative (for $\mathcal{H}_{0}$), and the second term corresponds to the mean time from then on till it crosses the threshold. Using the Gaussian approximation of Lemma $2$, the $t_{k}$'s are the order statistics of i.i.d. Gaussian random variables and hence, the $\tilde{F}_{k}$'s can be computed. (See, for example, \cite{gaussiannonidentical}). A similar approximation can be written for $\mathbb{E}_{1}[N]$.
\par Next, we compute approximate expressions for $P_{FA}$ and $P_{MD}$.
\par Under the same setup of large $\gamma_{0l}, \gamma_{1l}, \beta_0, \beta_1$, for $P_{FA}$ analysis, we assume that all local nodes are making correct decisions. Then for false alarm, the dominant event is $\{N^{1}<t_{1}\}$. Also, for reasonable performance, $P_{0}(N^{0}<t_{1})$ should be small. Then, the probability of false alarm, $P_{FA}$, can be approximated as
\begin{align*}
P_{FA} = P_{0}(N^{1}<N^{0}) &\geq P_{0}(N^{1}<t_{1},N^{0}>t_{1})
\end{align*}
\begin{align}
& \approx P_{0}(N^{1}<t_{1}).
\label{eqn:pfa_approx}
\end{align}
Also, 
\begin{align*}
P_{0}(N^{1}<N^{0})&\leq P_{0}(N^{1}<\infty)
\end{align*}
\begin{align}
&=P_{0}(N^{1}<t_{1})+P_{0}(t_{1}\leq N^{1}<t_{2})+\cdots
\label{eqn:P_expand}
\end{align}

The first term in the RHS of (\hspace{-0.15cm}~\ref{eqn:P_expand}) should be the dominant term since after $t_{1}$, the drift of $F_{k}$ will have the desired sign (will at least be in the favourable direction) with a high probability.
\par Equations (\hspace{-0.15cm}~\ref{eqn:pfa_approx}) and (\hspace{-0.15cm}~\ref{eqn:P_expand}) suggest that $P_{0}(N^{1}<t_{1})$ should serve as a good approximation for $P_{FA}$. Similar arguments show that $P_{1}(N^{0}<t_{1})$ should serve as a good approximation for $P_{MD}$. In the following, we provide approximations for these. 
\par Let $\hat{Z}_k$ before $t_{1}$ have mean 0 and probability distribution symmetric about $0$. This will happen if $\mathbb{E}[Z_k] = 0$, distribution of $Z_k$ is symmetric about $0$ and $\mu_0 + \mu_1 = 0$. Then, from the Markov property of the random walk $\{W_{k}\}$, before $t_{1}$,
\begin{align*}
&P_{0}(N^{1}<t_{1})\approx
\end{align*}
\begin{align*}
\sum_{k=1}^{\infty} &P_{0}[\{W_{k}\geq -\log c\}\\
&\bigcap_{n=1}^{k-1}\{W_{n} < -\log c\}|t_{1}>k]P_{0}(t_{1}>k)\\
=\sum_{k=1}^{\infty} &P_{0}[\{W_{k}\geq -\log c\}|\bigcap_{n=1}^{k-1}\{W_{n} < -\log c\}]\\
&P_{0}[\bigcap_{n=1}^{k-1}\{W_{n} < -\log c\}]P_{0}(t_{1}>k)\\
=\sum_{k=1}^{\infty} &P_{0}[W_{k}\geq -\log c|(W_{k-1} < -\log c)]\\
&P_{0}(\sup_{1 \leq n \leq k-1}W_{n} < -\log c)[1-\Phi_{t_{1}}(k)]\\
=\sum_{k=1}^{\infty} &[\int_{u=0}^{\infty} P_{0}(\hat{Z}_{k}>u){f}_{W_{k-1}}(-\log c - u)du]\\
&P_{0}(\sup_{1\leq n\leq k-1}W_{n} < -\log c).[1-\Phi_{t_{1}}(k)], 
\end{align*}
where $\Phi_{t_{1}}$ is the CDF of $t_{1}$. We can find a lower bound to the above expression by using\\ $P_{0}(\displaystyle \sup_{1\leq n\leq k-1}W_{n} < -\log c) \geq 1-2P_{0}(F_{k-1}\geq-\log c)$\\ \\(\cite{billingsley}, page $525$) and an upper bound by replacing $\displaystyle \sup_{1\leq n\leq k-1}W_{n}$ by $W_{k-1}$. 
\par Similarly, $P_{MD}$ can be approximated as\\
\begin{align*}
P_{MD}\gtrsim \sum_{k=1}^{\infty} &[\int_{u=0}^{\infty} P_{1}(\hat{Z}_{k}<-u){f}_{W_{k-1}}(\log \beta + u)du]\\
&[1-2P_{1}(W_{k-1} \leq \log \beta)][1-\Phi_{t_{1}}(k)],
\end{align*}
and 
\begin{align*}
P_{MD}\lesssim \sum_{k=1}^{\infty} &[\int_{u=0}^{\infty} P_{1}(\hat{Z}_{k}<-u){f}_{W_{k-1}}(\log \beta + u)du]\\
&P_{1}(W_{k-1} > \log \beta)[1-\Phi_{t_{1}}(k)].\\ 
\end{align*}
In the above expressions, $\mbox{f}_{W_{k-1}}$ stands for the probability density function of $W_{k-1}$.

\begin{figure}[]
                \centering
               \includegraphics[width=0.5\textwidth,height=5cm]{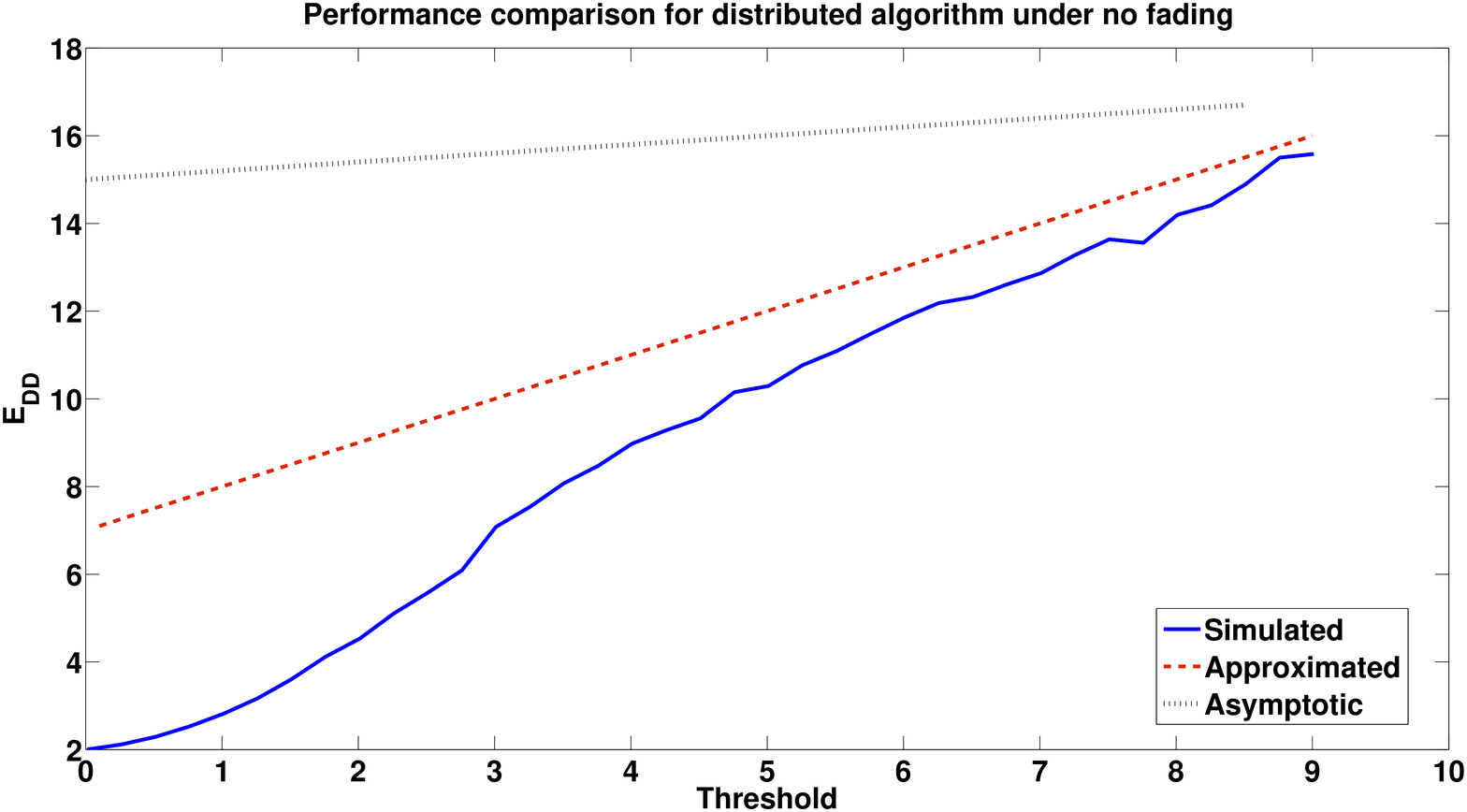}
\caption{Performance of distributed algorithm under Gaussian noise at the FC}
\label{fig:approx_compare_nofading}
\end{figure}

\begin{figure}[]
                \centering
               \includegraphics[width=0.5\textwidth,height=5cm]{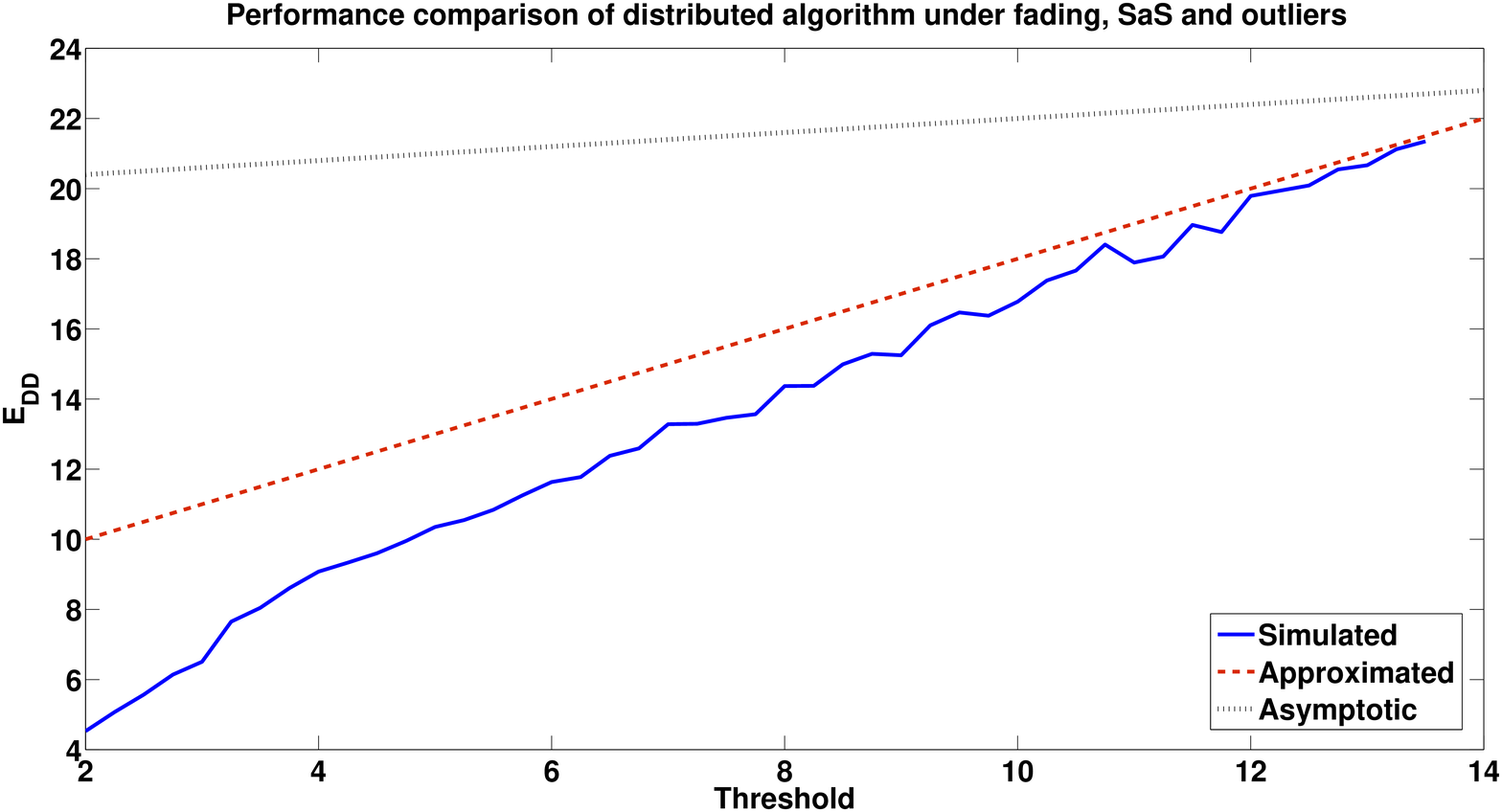}
\caption{Performance of distributed algorithm under Gaussian and S$\alpha$S noise, $5\%$ outliers and block Log $\mathcal{N}$ shadowing - Rayleigh fast fading (with outliers at local nodes).}
\label{fig:approx_compare}
\end{figure}

\begin{figure}[]
                \centering
               \includegraphics[width=0.5\textwidth,height=5cm]{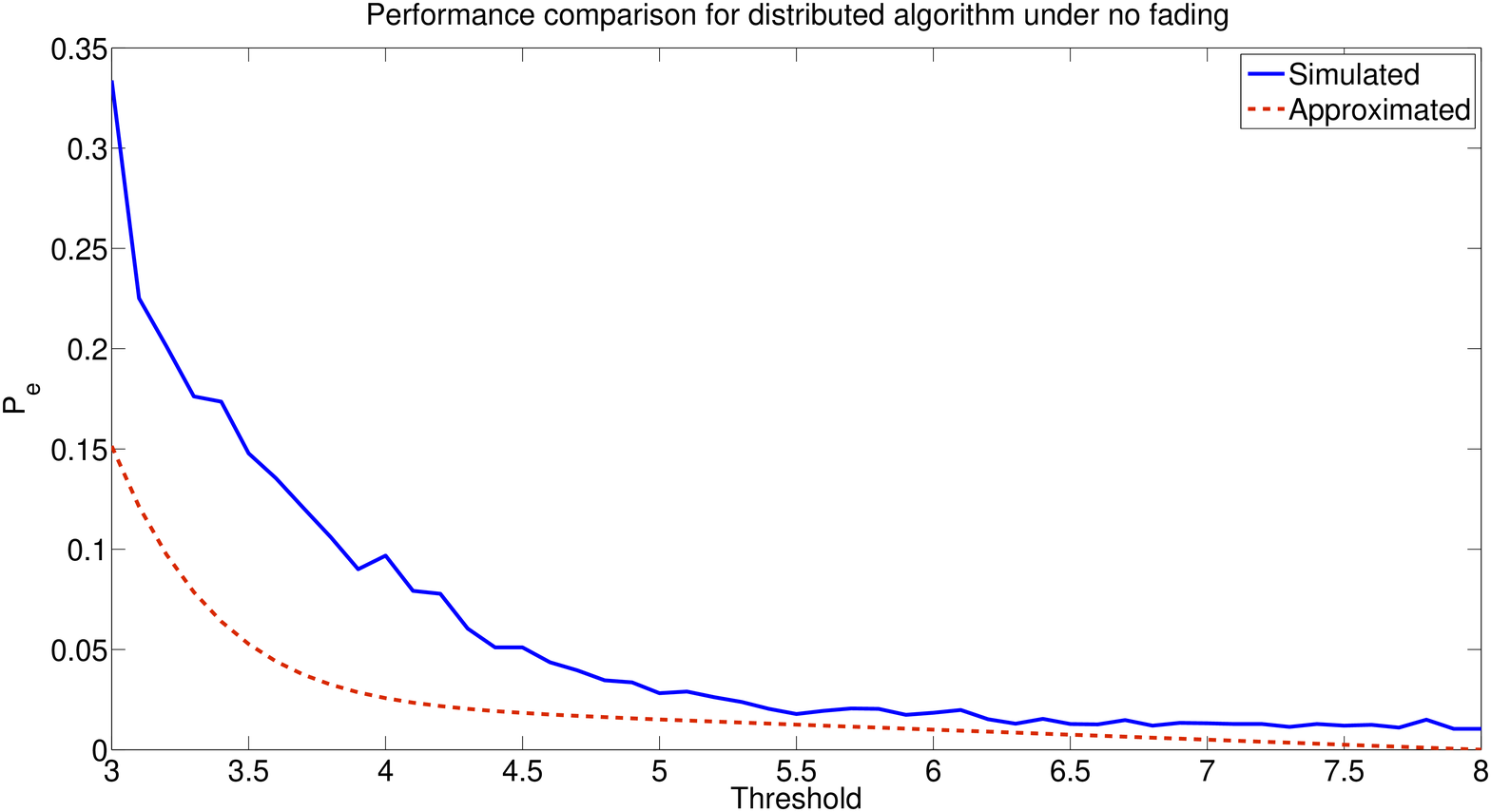}
\caption{Performance of distributed algorithm under Gaussian noise at the FC}
\label{fig:pe_approx_compare_nofading}
\end{figure}

\begin{figure}[]
                \centering
               \includegraphics[width=0.5\textwidth,height=5cm]{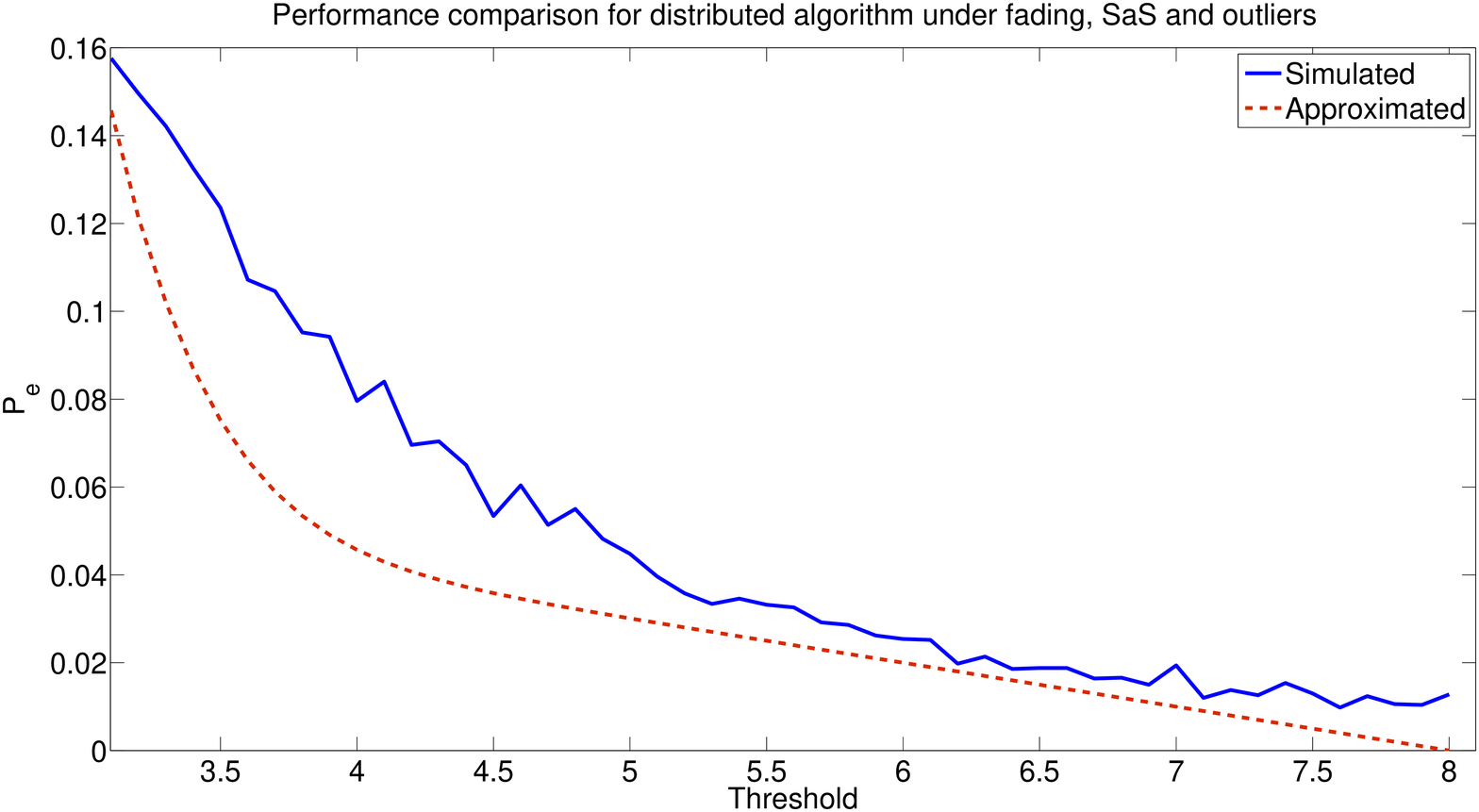}
\caption{Performance of distributed algorithm under Gaussian and S$\alpha$S noise, $5\%$ outliers and block Log $\mathcal{N}$ shadowing - Rayleigh fast fading (with outliers at local nodes).}
\label{fig:pe_approx_compare}
\end{figure}

Figures~\ref{fig:approx_compare_nofading} and~\ref{fig:approx_compare} show the comparison of simulation, approximation and asymptotics for $\mathbb{E}_i[N]$. (Please see Section~\ref{section:distri simulation} for details on the simulation setup). Figure~\ref{fig:approx_compare_nofading} shows the results when there is no fading and Figure~\ref{fig:approx_compare} shows the case wherein there is fading, EMI and outliers. We see that the approximation explains the simulation results much better than the asymptotics. We also get approximation for $P_e$ and see (in Figures~\ref{fig:pe_approx_compare_nofading} and~\ref{fig:pe_approx_compare}) that the approximations are close to the simulation results for small $P_e$.

\section{Simulation results for Distributed algorithm}

\begin{figure}[h]
        \centering
        \begin{subfigure}[b]{0.5\textwidth}
                \centering
                \includegraphics[width=\textwidth,height=5cm]{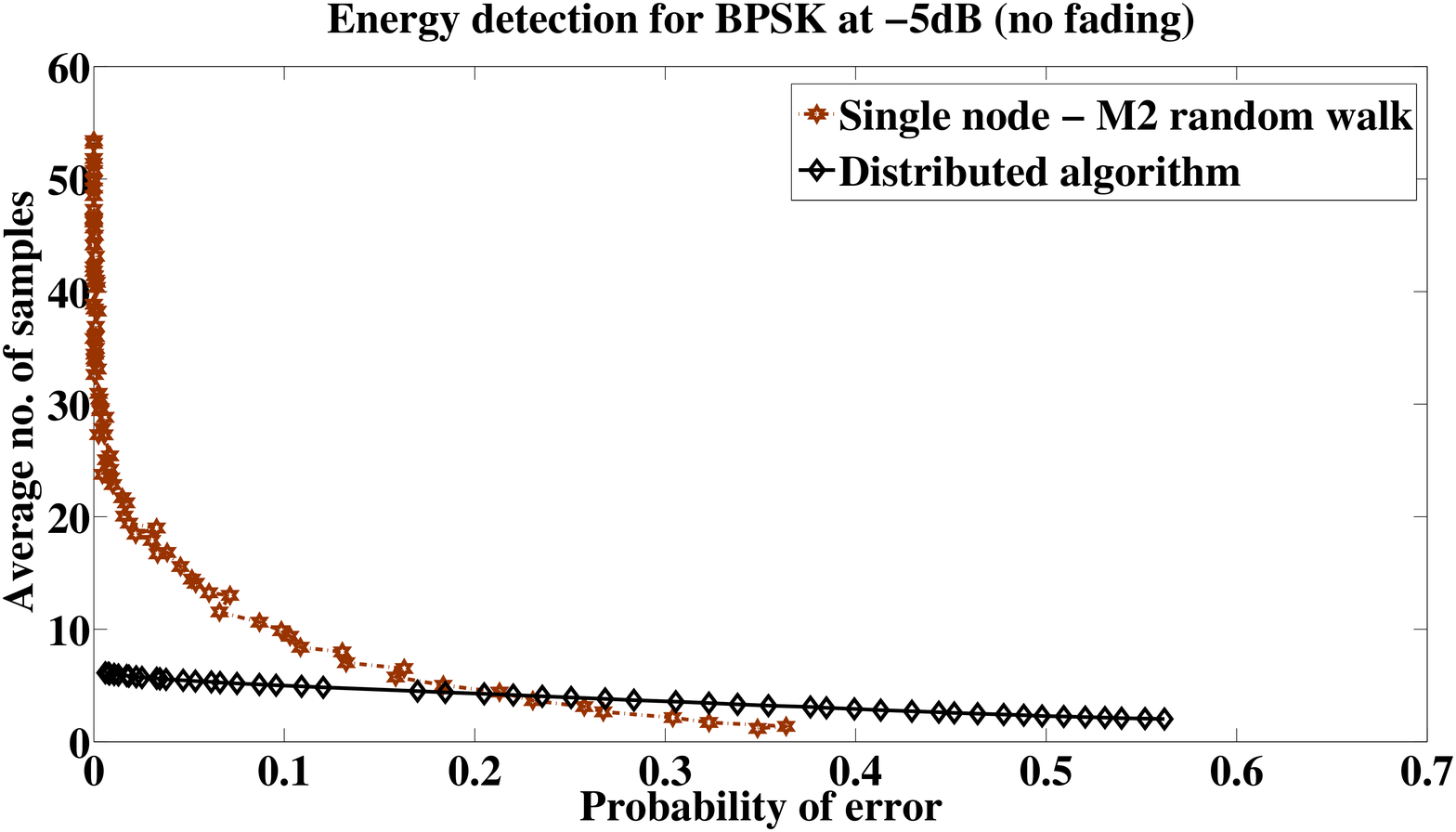}
        \end{subfigure}
        \begin{subfigure}[b]{0.5\textwidth}
                \centering
                \includegraphics[width=\textwidth,height=5cm]{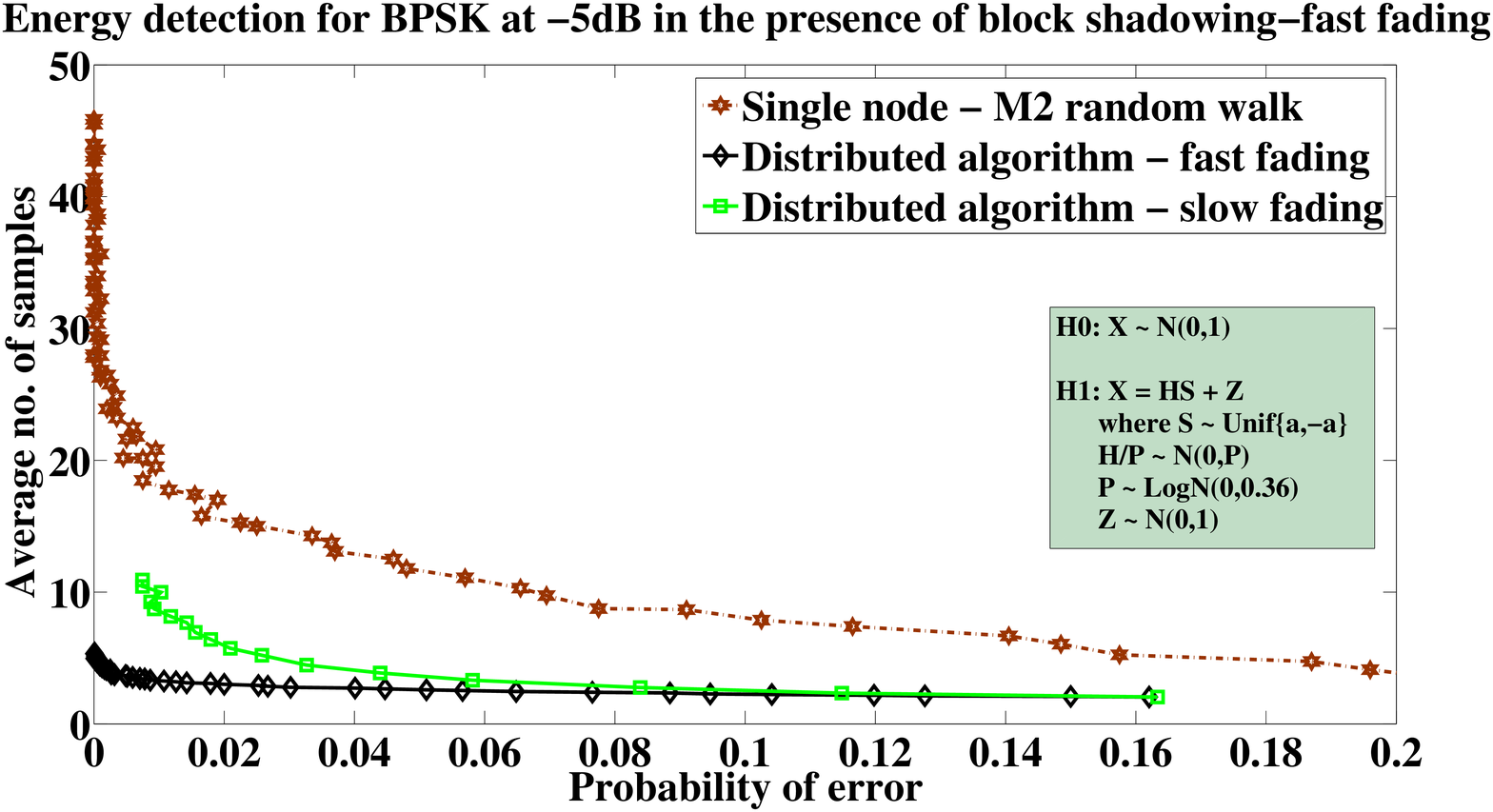}
         \end{subfigure}
         \caption{Energy detection in the presence of Gaussian noise. Top: Without fading. Bottom: Log $\mathcal{N}$ shadowing - Rayleigh fast fading.}
\label{fig:distri_gauss}
\end{figure}
\begin{figure}[h]
        \centering
        \begin{subfigure}[b]{0.45\textwidth}
                \includegraphics[width=\textwidth,height=5cm]{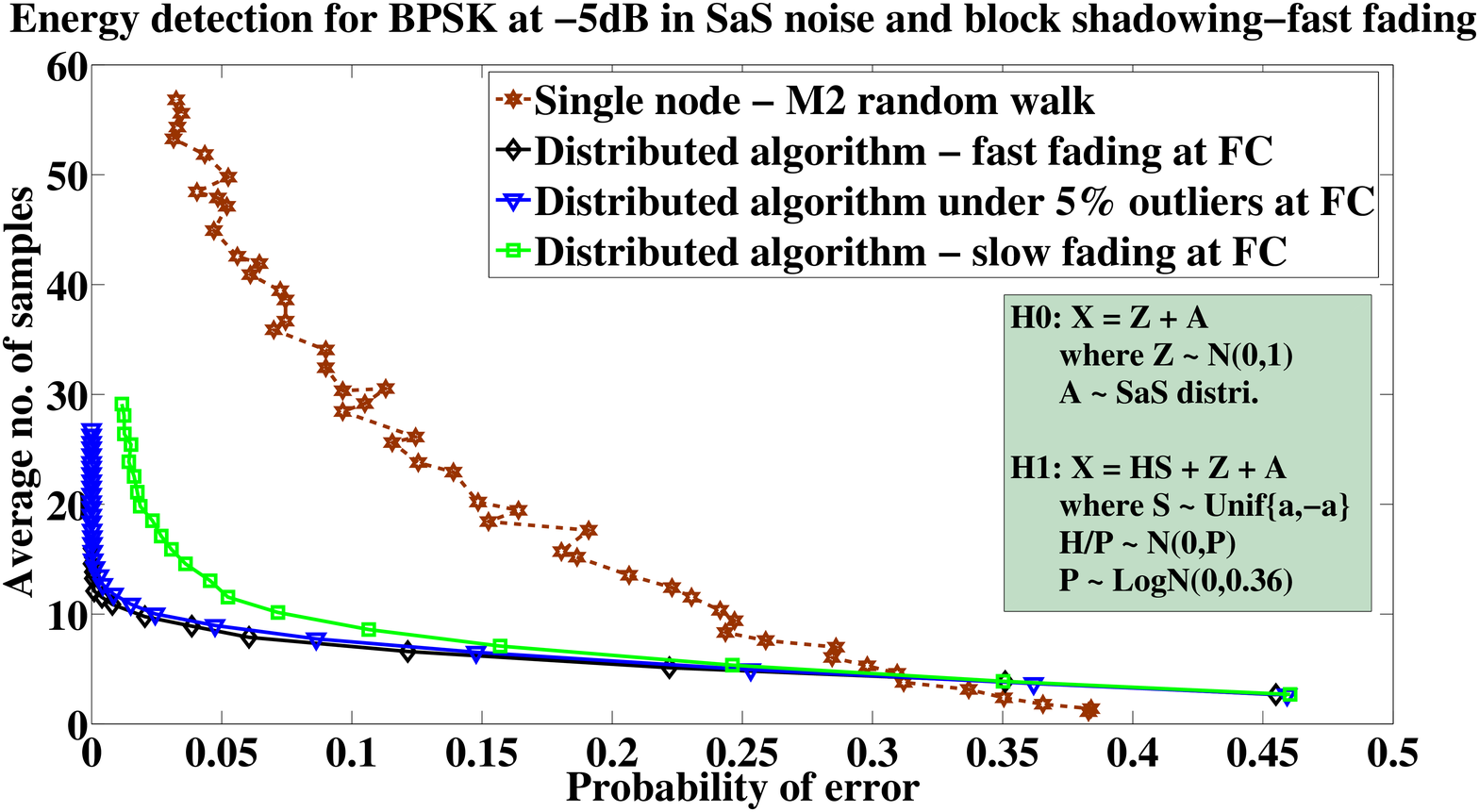}
		 \end{subfigure}
        \begin{subfigure}[b]{0.45\textwidth}
                \centering
        \centering
                \includegraphics[width=\textwidth,height=5cm]{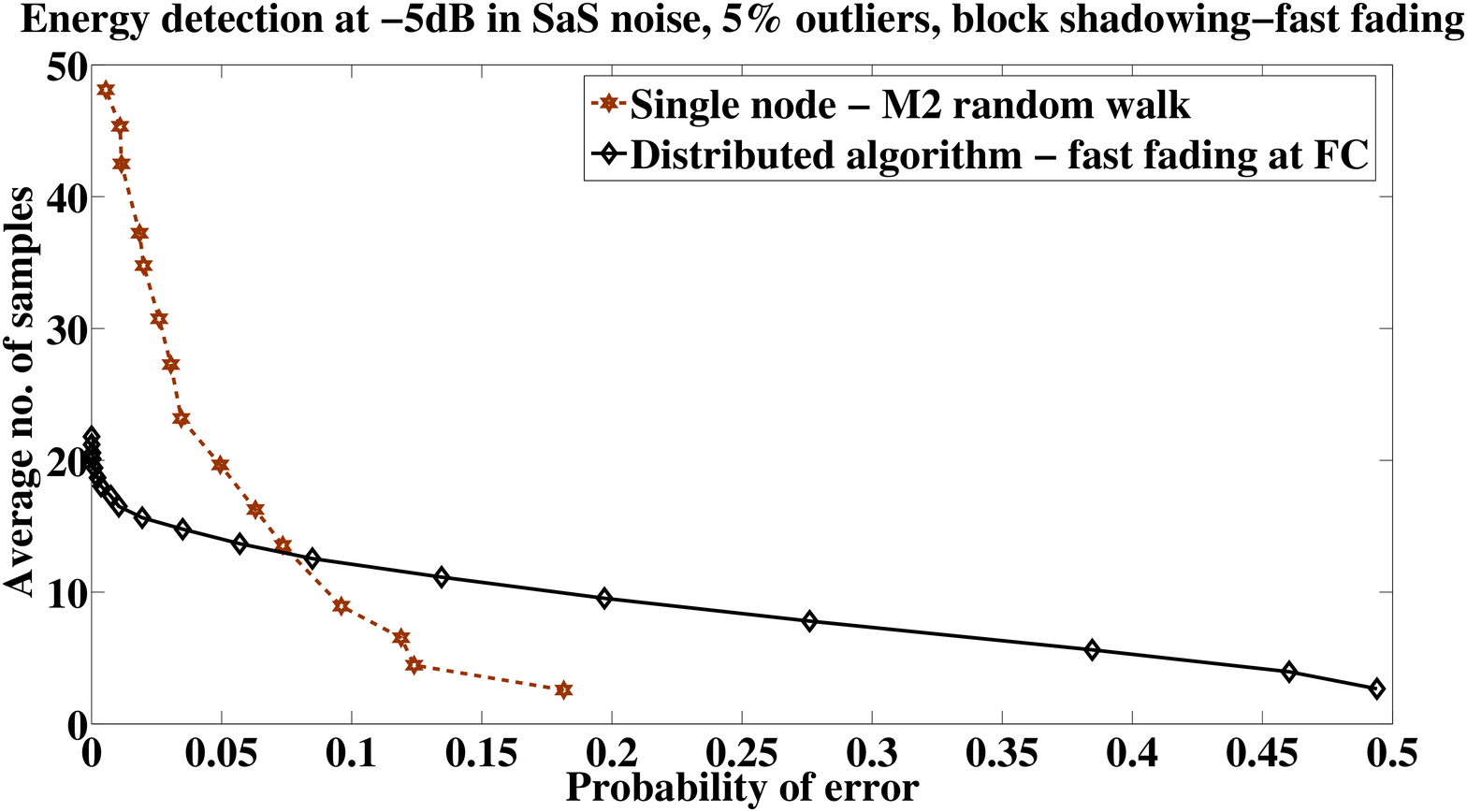}
         \end{subfigure}
\caption{Energy detection in the presence of Gaussian and S$\alpha$S noise, $5\%$ outliers and block Log $\mathcal{N}$ shadowing - Rayleigh fast fading. Top: Without outliers at local nodes. Bottom: With outliers at local nodes.}
\label{fig:distri_SaS_fading}
\end{figure}

We have considered $L=5$ local nodes reporting their decisions to the FC. The distributions of fading, EMI and outliers at the local nodes and the FC are the same as in Section~\ref{section:survey}. Also, $b_0=1, b_1=1$. The receiver noise at the local nodes is $\mathcal{N}(0,1)$ and at the FC is $\mathcal{N}(0,5)$. From Figures~\ref{fig:distri_gauss} and~\ref{fig:distri_SaS_fading}, we see that the distributed algorithm performs much better than the single node algorithm using $M^2$-random walk, especially in the low probability of error regime. Figure~\ref{fig:distri_gauss} shows the comparison when the local nodes run $M^2$-random walk and FC runs $M$-random walk in the presence of Gaussian noise and fading. From Figure~\ref{fig:distri_SaS_fading} we see that the distributed algorithm performs better in the presence of EMI (along with shadowing-fading) at the local nodes and the FC. It also shows that the presence of $5\%$ outliers (along with shadowing-fast fading) at the FC does not make a considerable difference in the performance of the distributed algorithm. We have considered the presence of outliers only when the signal is present (under $\mathcal{H}_1$ at local nodes and under both the hypotheses at the FC). With all these impairments, the reporting channel becomes bad and the improvement over the single node case is seen for small probability of error only (Figure~\ref{fig:distri_SaS_fading} bottom).
\par We see that the distributed algorithm performs much better than the single node $M^2$-random walk, especially at low probability of error. It is also quite robust to the effects of fading, EMI and outliers at the local nodes and the FC.

\label{section:distri simulation}
\section{Conclusion}
\par We propose a distributed algorithm for spectrum sensing in Cognitive radio. Various impairments such as additive noise, EMI, shadowing and multipath effects and outliers were taken into account while designing the algorithm. The local nodes perform energy detection for lack of knowledge about the primary's transmission parameters and the FC is signalled via BPSK. We find that robust versions of random walk algorithm developed recently, perform well in case of energy detection at the local nodes and binary signalling over the reporting MAC channel. We have performed simulations to demonstrate this and have theoretically validated the observations.
\label{section:conclusion}

\section{Appendix}
\par The proofs of Theorems $1-4$ are provided in this appendix. We will use the following notation:
\begin{align*}
&N_l^0 = \inf\{n: T_{nl} \le -\gamma_{0l}\},\\
&N_l^1 = \inf\{n: T_{nl} \ge \gamma_{1l}\},\\
&\tau_l(-\gamma_{0l}) = \text{the last time random walk }T_{nl} \text{ is above}-\gamma_{0l},\\
&\tau = \max_{l} \tau_l(-\gamma_{0l}),\\
&\nu(a) = \text{Starting from } 0, \text{ the first time }W_k \text{ crosses }a \\
&\text{ \hspace{1cm}when all local nodes are transmitting }-b_0,\\
&\tau(c) = \max_l \tau_l(-\gamma_l |\log c|).
\end{align*}

\textbf{Proof of Theorem $1$}. We show $P_0[N < \infty]=1$. Similarly we can show for $P_1[N < \infty]=1$. 
\par Under $\mathcal{H}_0$, $T_{nl}$ is a random walk with finite negative mean, for each $l$. Thus, $T_{nl} \rightarrow -\infty$ a.s. and hence $\tau_l(-\gamma_{0l}) < \infty$ a.s. for any finite $\gamma_{0l}$. Therefore $\tau < \infty$ a.s. After $\tau$, all local nodes transmit $-b_0$ and hence increments of $W_k$ have a negative mean ($=\Delta_0$). Therefore,
\begin{equation}
N < \tau + \nu (-W_{\tau+1} - \beta_0).
\label{equation:appendix_eqn1}
\end{equation}
Since $\tau < \infty$ a.s. and $\mathbb{E}[|Y_k|] < \infty$, $|W_{\tau+1}| < \infty$ a.s. and hence $\nu (-W_{\tau+1} - \beta_0) < \infty$ a.s. Therefore, $P_0[N < \infty] = 1$. \hfill $\blacksquare$\\

\textbf{Proof of Theorem $2$}. We prove for $\mathcal{H}_0$. From ($\hspace{-0.12cm}~\ref{equation:appendix_eqn1}$),
\begin{equation*}
\frac{N}{|\log c|} \le \frac{\tau(c)}{|\log c|} + \frac{\nu(-W_{\tau(c) + 1} - |\log c|)}{|\log c|}.
\end{equation*}
\par Also, from \cite{gut}
\begin{equation*}
\lim_{c \rightarrow 0} \frac{\tau(c)}{|\log c|} = \frac{1}{D_{tot}^0} \text{a.s.}
\end{equation*}
Furthermore,
\begin{align*}
\frac{\nu (-|\log c| - W_{\tau(c)+1})}{|\log c|}& \le \frac{\nu(-|\log c|)}{|\log c|} \\
&+ \frac{\nu(-W_{\tau(c)+1})}{W_{\tau(c)+1}} \frac{W_{\tau(c)+1}}{\tau(c)+1} \frac{\tau(c)+1}{|\log c|}
\end{align*}
and from \cite{gut},
\begin{equation*}
\lim_{c \rightarrow 0} \frac{\nu(-|\log c|)}{|\log c|} = -\frac{1}{\Delta_0} \text{ a.s.}
\end{equation*}
Also, by Strong Law of Large Numbers (SLLN) (since $\tau(c) \to \infty$ a.s. as $c \downarrow 0$)
\begin{equation*}
\lim_{c \downarrow 0} \frac{W_{\tau(c)+1}}{\tau(c)+1} \le \mathbb{E}[\xi_1^*] \text{ a.s.}
\end{equation*}
Thus, 
\begin{equation*}
\lim_{c \rightarrow 0} \frac{\nu(-W_{\tau(c)+1})}{W_{\tau(c)+1}} \frac{W_{\tau(c)+1}}{\tau(c)+1} \frac{\tau(c)}{|\log c|} \le \Big(\frac{-1}{\Delta_0}\Big) \frac{\mathbb{E}[|\zeta_1^*|]}{D_{tot}^0} \text{ a.s.}
\end{equation*}\hfill $\blacksquare$\\

\textbf{Proof of Theorem $3$}. We prove the result for $P_{FA}$. It holds for $P_{MD}$ in the same way. 
\par We have,
\begin{equation*}
P_{FA} = P_0[\text{Declare } \mathcal{H}_1 \text{ upto } \tau(c)] + P_0[\text{Declare } \mathcal{H}_1 \text{ after } \tau(c)].
\end{equation*}
The first term on the $RHS$,
\begin{align*}
&P_0\Big[\text{Declare } \mathcal{H}_1 \text{ upto } \tau(c)\Big] \\
&\le P_0\Big[\cup_l\{l^{th} \text{ random walk crosses }\gamma_l|\log c| \text{ upto } \tau(c)\}\Big]\\
&+P_0\Big[\hat{W}_0=0, \hat{W}_k \text{ crosses } |\log c| \text{ upto } \tau(c) \\
&\text{ with }\hat{W}_k=\sum_{n=1}^k \xi_n^*\Big]
\end{align*}
\begin{align}
&\le \sum_l P_0\Big[\sup_{0 \le k \le \tau(c)} T_{kl} > \gamma_l |\log c|\Big] \\
&+ P_0\Big[\sup_{0 \le k \le \tau(c)} \hat{W}_k > |\log c|\Big].
\label{eqn:justabove}
\end{align}
\par The first term on the RHS in (\hspace{-0.15cm}~\ref{eqn:justabove}),
\begin{align*}
P_0\Big[\sup_{0 \le k \le \tau(c)} T_{kl} > \gamma_l |\log c|\Big] & \le P\Big[M_{0l} > \gamma_l |\log c|\Big]\\
& \le e^{-\Gamma_{0l} \gamma_l |\log c|},
\end{align*}
where $M_{0l} = \sup_{k \ge 0} T_{kl}$ and $\mathbb{E}_0[e^{\Gamma_{0l}\hat{X}_{1l}}]=1$. Since $\Gamma_{0l} > 0$, this goes down exponentially to $0$ with $|\log c|\rightarrow\infty$ at rate $\Gamma_{0l}\gamma_l$.
\par Now consider $P_0[\sup_{0 \le k \le \tau(c)} \hat{W}_k > |\log c|]$, the second term on the RHS of (\hspace{-0.12cm}~\ref{eqn:justabove}). From Theorem $10$ in \cite{zolotarev}, since $\{\hat{W}_k\}$ is a submartingale, we get, for an $\alpha > 0$ with $\mathbb{E}[e^{\alpha \xi_1^*}] < \infty$ (by taking function $g(x) = e^{\alpha x}-1, \alpha > 0$ in Theorem $10$ of \cite{zolotarev}),
\begin{equation*}
P[\sup_{k \le n} \hat{W}_k \ge x] \le \frac{\mathbb{E}[e^{\alpha\hat{W}_n}]-1}{e^{\alpha x}-1}.
\end{equation*}
Therefore, since $\tau(c)$ is independent of $\{\hat{W}_k\}$,
\begin{align*}
&\sum_{n=1}^\infty P[\sup_{0 \le k \le n} \hat{W}_k \ge |\log c|]P[\tau(c)=n] \\
&\le \sum_{n=1}^\infty  P[\tau(c)=n]\Big(\frac{\phi_{\xi^*}(\alpha_0)^n - 1}{e^{\alpha_0|\log c|}-1}\Big)\\
&=(e^{\alpha_0|\log c|}-1)^{-1} \mathbb{E}[e^{\tau(c)\log \phi_{\xi^*}(\alpha_0)}-1]\\
\end{align*}
\vspace{-1cm}
\begin{align}
&\le (e^{\alpha_0|\log c|}-1)^{-1} \mathbb{E}[(e^{\eta\tau(c)}-1)].
\label{equation:theorem3eq1}
\end{align}
if $\log \phi_{\xi^*}(\alpha_0) \le \eta$. Thus, we have from Lemma $3$ below, exponential decay if $k_2 < \alpha_0$.
\par Next consider, for an $r, 0<r<1$, 
\begin{align*}
&P_0\Big[\text{Declare } \mathcal{H}_1 \text{ after } \tau(c)\Big] \\
&\le P_0\Big[\text{Declare } \mathcal{H}_1 \text{ after } \tau(c) \text{ if } W_{\tau(c)+1} \le r |\log c|\Big]\\
&+P_0\Big[W_{\tau(c)+1} > r |\log c|\Big].
\end{align*}
Since,
\begin{equation*}
P_0[W_{\tau(c)+1} > r |\log c|] \le P[\sup_{0 \le k \le \tau(c)+1} \hat{W}_k > r |\log c|],
\end{equation*}
from ($\hspace{-0.15cm}~\ref{equation:theorem3eq1}$) and Lemma $3$,
\begin{equation*}
P_0[W_{\tau(c)+1} > r |\log c|] \le k_1' \frac{\mathbb{E}[e^{\eta(\tau(c)+1)}]-1}{k_1'e^{r\alpha_0 |\log c|}-1} \rightarrow 0 
\end{equation*}
exponentially in $|\log c|$ if $r\alpha_0 > k_2$ for some $0<r<1$.
\par Next consider
\begin{align*}
&P_0\Big[\text{Declare } \mathcal{H}_1 \text{ after } \tau(c) \text{ if } W_{\tau(c)+1} \le r |\log c|\Big]\\
&\le P_0\Big[\text{Declare } \mathcal{H}_1 \text{ after } \tau(c) \text{ if } W_{\tau(c)+1} = r |\log c|\Big]\\
&\le P\Big[\text{random walk starting with zero } \\
&\text{ and increments } Z_k-Lb_0 \text{ has max} > (1-r)|\log c|\Big]\\
&\le e^{-\Gamma_0(1-r)|\log c|} \rightarrow 0,
\end{align*}
exponentially because $\Gamma_0 > 0$ where $\mathbb{E}[e^{\Gamma_0(Z_1 - Lb_0)}] = 1$.
Thus,
\begin{align*}
P_{FA} \le &k_1 e^{-\alpha_0|\log c|} e^{k_2|\log c|}+k_1' e^{-\alpha_0 r |\log c|}e^{k_2 |\log c|}\\
&+e^{-\Gamma_0(1-r)|\log c|}+\sum_{l=1}^Le^{-\Gamma_{0l}\gamma_l|\log c|}
\end{align*}
and $\lim_{c \downarrow 0} \frac{P_{FA}}{c^{r'}} < \infty$ where $r' < \min\{\alpha_0r - k_2, \Gamma_0(1-r), \Gamma_{0l}\gamma_{l}, l=1,..., L\}$.\hfill $\blacksquare$\\

\textbf{Lemma $3$}. Let there be an $\eta$ such that $R_0 > \eta > 0$ and $\gamma_l'$ is the smallest positive constant with $\phi_{il}(\gamma_l')= e^{-\eta}$ for all $l=1,..., L$. Also, there is $\alpha_1 >0$ such that 
$\bar{\phi}_{il}(\alpha_1) \triangleq \mathbb{E}[e^{-\alpha_1\hat{X}_{1l}}] < \infty$ for all $l=1,..., L$ and $\eta + \log \bar{\phi}_{0l}(\alpha_1) < 0$. Then, $\limsup_{c \downarrow 0} \frac{\mathbb{E}[e^{\eta \tau(c)}]}{k_1 e^{k_2|\log c|}} \le 1$ when $k_1$ is a constant and $k_2 = \sum_l \gamma_l \gamma_l'$.\\

\textbf{Proof.} For any $c$, for $l=1,...,L$
\begin{align*}
\mathbb{E}[e^{\eta\tau_l(c)}] &= \sum_{n=0}^\infty e^{\eta n}P[\tau_l(c)=n]\\
&\le \sum_{n=0}^{\infty} e^{\eta n}P(T_{(n+1)l} \le -\gamma_l |\log c|)\\
&= e^{-\eta}\sum_{n=1}^{\infty} e^{\eta n}P(-T_{nl} \ge \gamma_l |\log c|)
\end{align*}
\vspace{-.5cm}
\begin{align}
\hspace{1.4cm}&\le \sum_{n=1}^\infty e^{\eta n} \frac{\bar{\phi}_{il}(\alpha_1)^n}{e^{\gamma_l|\log c|}},
\label{eqn:lemmamy2}
\end{align}
by Markov inequality. Also, RHS of $(\hspace{-0.18cm}~\ref{eqn:lemmamy2})$ is finite if $\eta + \log \bar{\phi}_{0l}(\alpha_1) < 0$.
Thus, under our assumptions, from \cite{meiners}, there exist positive $\gamma_l'$ such that
\begin{equation*}
\lim_{c \downarrow 0} \frac{\mathbb{E}[e^{\eta \tau_l(c)}]}{f_l(\gamma_l)e^{\gamma_l |\log c| \gamma_l'}} = 1 \text{ for each }l,
\end{equation*}
when $f_l(\gamma_l)$ is a constant provided in \cite{meiners}.
Thus, since $\tau_1(c),\tau_2(c),..., \tau_L(c),$ are independent,
\begin{align*}
\mathbb{E}[e^{\eta \tau(c)}] &\le \mathbb{E}[e^{\eta \sum_{l=1}^L \tau_l(c)}]= \prod_{l=1}^L\mathbb{E}[e^{\eta \tau_l(c)}].
\end{align*}
Therefore,
\begin{equation*}
\lim_{c \downarrow 0} \frac{\mathbb{E}[e^{\eta \tau(c)}]}{\prod_{l=1}^Lf_l(\gamma_l)e^{|\log c| \sum_l r_l \gamma_l'}} \le 1.
\end{equation*}\hfill $\blacksquare$

\textbf{Proof of Theorem $4$}. Define
\begin{align*}
A_l = \{\text{local node } l \text{ makes a wrong decision some time}\}.
\end{align*}
Then,
\begin{align*}
P_{FA} &= P_0(\text{ declare }\mathcal{H}_1) \\
&\le P_0(\text{ declare }\mathcal{H}_1|\cap_l A_l^C)P(\cap_l A_l^C) + \sum_{l=1}^LP_0(A_l).
\end{align*}
From the text below equation ($\hspace{-0.12cm}~\ref{equation:4}$),
\begin{align*}
P_0[A_l] \le P_0[\sup_{n \ge 0}T_{nl} &\ge \rho_l |\log c|]\\
&\sim f_l(\rho_l |\log c|)(\rho_l |\log c|)^{-r_1}
\end{align*}
where $f_l$ is a slowly varying function. Thus,
%\begin{equation}
%P_0[\cup_l A_l] \le \sum_l P(A_l)\\
%\precsim \Big( \sum_l \frac{f_l(\rho_l |\log c|)}{\rho_l^{r_1}}\Big)\frac{1}{(|\log c|)^{r_1}} 
%\label{equation:appendix_eqn3}
%\end{equation}
%where $f(x) \precsim g(x)$ denotes $\limsup_{x \to \infty} \frac{f(x)}{g(x)} \le constant$.
\begin{align*}
P_0(\text{Declare }\mathcal{H}_1) &\precsim P_0(\cap_l A_l^c)P_0(\text{Declare }\mathcal{H}_1 | \cap_l A_l^c)
\end{align*}
\begin{align}
&&\Big( \sum_l \frac{f_l(\rho_l |\log c|)}{\rho_l^{r_1}}\Big)\frac{1}{(|\log c|)^{r_1}}.
\label{equation:appendix_eqn3}
\end{align} 
Consider
\begin{align}
\begin{split}
P_0(\text{Declare }\mathcal{H}_1 |\cap_l A_l^c) &= P_0(\text{Declare }\mathcal{H}_1 \text{upto } \tau(c)| \cap_l A_l^c)\\
&+P_0(\text{Declare }\mathcal{H}_1 \text{after } \tau(c)| \cap_l A_l^c)\\
&\le P_0(\sup_{0 \le k \le \tau(c)} \sum_{k=1}^n Z_k > |\log c|)\\
&+P_0(\text{Declare }\mathcal{H}_1 \text{after } \tau(c)| \cap_l A_l^c)
\label{equation:appendix_eqn4}
\end{split}
\end{align}
\par Also, from \cite{foss}, since $Z_k$ distribution $\in R(-r_2 -1)$,
\begin{align*}
&P_0[\sup_{1 \le k \le \tau(c)} \sum_{n=1}^k Z_n > |\log c|] \\
&\sim \mathbb{E}[\tau(c)]g_1(|\log c|)|\log c|^{-r_2-1}
\end{align*}
\begin{align}
&\sim |\log c|^{-r_2}g_1(|\log c|).
\label{equation:appendix_eqn5}
\end{align}
where $g_1$ is a slowly varying function.\\

\par The second term on $RHS$ of ($\hspace{-0.15cm}~\ref{equation:appendix_eqn4}$), for a $\delta, 0 < \delta <1$,
\begin{align*}
&P_0(\text{Declare }\mathcal{H}_1 \text{ after } \tau(c)| \cap_l A_l^c)\\
&\le P[\text{random walk at FC with mean } \Delta_0\\
&\text{ and initial condition }\sum_{k=1}^{\tau(c)+1} Z_k \text{ crosses } |\log c|]\\
&\le P[\text{FC random walk with mean } \Delta_0 \text{ and initial condition }\\
& \delta |\log c| \text{ crosses } |\log c|] + P[\sum_{k=1}^{\tau(c)+1} Z_k > (1-\delta)|\log c|].
\end{align*}
Also, for slowly varying functions $g_2, g_3$,
\begin{align*}
&P[\sum_{k=1}^{\tau(c)+1} Z_k > (1-\delta)|\log c|]\\
&\le P[\sup_{0 \le k \le \tau(c)+1} \sum_{j=0}^{k} Z_j > (1-\delta)|\log c|]\\
&\sim g_2((1-\delta)|\log c|)\mathbb{E}[\tau(c)+1] |\log c|^{-r_2-1}
\end{align*}
%\vspace{-1cm}
\begin{align}
&\sim g_2((1-\delta)|\log c|)|\log c|^{-r_2}
\label{equation:appendix_eqn6}
\end{align}
and \\
$P$[random walk with mean $-\Delta_0$ and initial condition $\delta |\log c|$ crosses $|\log c|$]
\begin{equation}
\le g_3((1-\delta)|\log c|)|\log c|^{-r_2}.
\label{equation:appendix_eqn7}
\end{equation}
From ($\hspace{-0.15cm}~\ref{equation:appendix_eqn3}$), ($\hspace{-0.15cm}~\ref{equation:appendix_eqn4}$), ($\hspace{-0.15cm}~\ref{equation:appendix_eqn5}$), ($\hspace{-0.15cm}~\ref{equation:appendix_eqn6}$) and ($\hspace{-0.15cm}~\ref{equation:appendix_eqn7}$),
\begin{align*}
P_{FA} &\precsim \Big( \sum_l \frac{f_l(\rho_l |\log c|)}{\rho_l^{r_1}}\Big)\frac{1}{(|\log c|)^{r_1}}\\ 
&+ g_1(|\log c|)|\log c|^{-r_2}) + g_2(1-\delta)|\log c|)|\log c|^{-r_2}\\
&+ g_3(1-\delta)|\log c|)|\log c|^{-r_2}. 
\end{align*}\hfill $\blacksquare$

\label{section:appendix}

\bibliography{references}		% expects file "myrefs.bib"

\begin{thebibliography}{10}

\bibitem{mitola}
J.~Mitola and J.~Maguire, G.Q., ``Cognitive radio: making software radios more
  personal,'' {\em IEEE Personal Communications}, vol.~6, no.~4, pp.~13--18,
  Aug 1999.

\bibitem{CR_survey}
I.~F. Akyildiz, B.~Lo, and R.~Balakrishnan, ``Cooperative spectrum sensing in
  cognitive radio networks: A survey,'' {\em Physical Communication}, vol.~4,
  no.~1, pp.~40--62, 2011.

\bibitem{CR_poor_survey}
E.~Axell, G.~Leus, E.~G. Larsson, and H.~V. Poor, ``Spectrum sensing for
  cognitive radio : State-of-the-art and recent advances,'' {\em Signal
  Processing Magazine, IEEE}, vol.~29, no.~3, pp.~101--116, May 2012.

\bibitem{govind}
Z.~Govindarajulu, {\em Sequential statistics}.
\newblock World Scientific Publishing Co., 2004.

\bibitem{low_snr_04}
A.~Sahai, N.~Hoven, and R.~Tandra, ``Some fundamental limits on cognitive
  radio,'' in {\em 42nd Allerton Conference on Communication, Control, and
  Computing}, 2004.

\bibitem{atapattu}
S.~Atapattu, C.~Tellambura, and H.~Jiang, {\em Energy Detection for Spectrum
  Sensing in Cognitive Radio}.
\newblock DOI 10.1007/978-1-4939-0494-5, SpringerBriefs in Computer Science,
  2014.

\bibitem{SalphaS_survey}
M.~Z. Win, P.~C. Pinto, and L.~A. Shepp, ``A mathematical theory of network
  interference and its applications,'' {\em Proceedings of the IEEE}, vol.~97,
  no.~2, pp.~205--230, 2009.

\bibitem{SalphaS_trans}
J.~Park, G.~Shevlyakov, and K.~Kim, ``Maximin distributed detection in the
  presence of impulsive alpha-stable noise,'' {\em IEEE Trans. on Wireless
  Communications}, pp.~1687 -- 1691, June 2011.

\bibitem{huber}
P.~J. Huber and E.~M. Ronchetti, {\em Robust Statistics}.
\newblock 2nd edition, A. John Wiley and sons Inc. publication, 2009.

\bibitem{tse}
D.~Tse and P.~Viswanath, {\em Fundamentals of Wireless Communication}.
\newblock Cambridge University press, 2005.

\bibitem{nakagami}
J.~D. Parsons, {\em The mobile radio propagation channel}.
\newblock 2nd edition, John Wiley and Sons, 2000.

\bibitem{logray}
F.~Hansen and F.~I. Meno, ``Mobile fading-rayleigh and lognormal
  superimposed,'' {\em IEEE Trans. on Vehicular Technology}, vol.~VT 26, no.~4,
  pp.~332--335, 1997.

\bibitem{ghas}
A.~Ghasemi and E.~Sousa, ``Spectrum sensing in cognitive radio networks:
  Requirements, challenges and design trade-offs,'' {\em IEEE Communications
  Magazine}, vol.~46, no.~4, pp.~32--39, 2008.

\bibitem{viswanath}
R.~Viswanathan and P.~K. Varshney, ``Distributed detection with multiple
  sensors i. fundamentals,'' {\em IEEE Proceedings}, vol.~85, pp.~54--63, 1997.

\bibitem{jointPHY}
G.~Feng, W.~Chen, and Z.~Cao, ``A joint phy-mac spectrum sensing algorithm
  exploiting sequential detection,'' {\em IEEE Signal Processing Letters},
  vol.~17, no.~8, pp.~703--706, Aug 2010.

\bibitem{latest}
F.~Lin, R.~C. Qiu, and J.~P. Browning, ``Spectrum sensing with small-sized data
  sets in cognitive radio: Algorithms and analysis,'' {\em IEEE Trans. on
  Vehicular Technology}, vol.~64, no.~1, pp.~77--87, Jan 2015.

\bibitem{hypt_survey}
Y.~Zeng, Y.~C. Liang, A.~T. Hoang, and R.~Zhang, ``A review on spectrum sensing
  for cognitive radio: Challenges and solutions,'' {\em EURASIP Journal on
  Advances in Signal Processing}, no.~Article id 381465, 2010.

\bibitem{fellouris}
G.~Fellouris and G.~V. Moustakides, ``Decentralized sequential hypothesis
  testing using asynchronous communication,'' {\em IEEE Trans. on Information
  Theory}, vol.~57, no.~1, pp.~534--548, 2011.

\bibitem{mei}
Y.~Mei, ``Asymptotic optimality theory for decentralized sequential hypothesis
  testing in sensor networks,'' {\em IEEE Trans. on Information Theory},
  vol.~54, no.~5, pp.~2072--2089, 2008.

\bibitem{uscslrt}
J.~K. Sreedharan and V.~Sharma, ``Nonparametric decentralized sequential
  detection via universal source coding,'' in {\em Information Theory and
  Applications Workshop (ITA)}, DOI 10.1109/ITA.2013.6502977, 2013.

\bibitem{entropy_icc}
S.~Ganguly, K.~R. Sahasranand, and V.~Sharma, ``A new algorithm for distributed
  nonparametric sequential detection,'' in {\em International Conference on
  Communications (ICC)}, 2014.

\bibitem{chamberland}
J.~Chamberland and V.~Veeravalli, ``Wireless sensors in distributed detection
  applications,'' {\em IEEE Signal Processing Magazine}, vol.~24, pp.~16--25,
  2007.

\bibitem{valli}
V.~Veeravalli, ``Sequential decision fusion: theory and applications,'' {\em
  Journal of the Franklin Institute}, vol.~336, pp.~301--322, 1999.

\bibitem{taposh}
T.~Banerjee, V.~Sharma, V.~Kavitha, and A.~Jayaprakasam, ``Generalized analysis
  of a distributed energy efficient algorithm for change detection,'' {\em IEEE
  Trans. on Wireless Communication}, vol.~10, pp.~91--101, 2011.

\bibitem{jithin_ncc}
K.~S. Jithin, V.~Sharma, and R.~Gopalarathnam, ``Cooperative distributed
  sequential spectrum sensing,'' in {\em National Conference on Communications
  (NCC)}, 2011.

\bibitem{quan}
Z.~Quan, S.~Cui, H.V.Poor, and A.~Sayed, ``Collaborative wideband sensing for
  cognitive radios,'' {\em IEEE Signal Processing Magazine}, vol.~25,
  pp.~60--73, 2008.

\bibitem{dualsprt}
J.~K. Sreedharan and V.~Sharma, ``Spectrum sensing using distributed sequential
  detection via noisy reporting mac,'' in {\em Signal Processing, Elsevier},
  vol.~106, pp.~159--173, January 2015.

\bibitem{liusensor}
K.~Liu and A.~M. Sayeed, ``Optimal distributed detection strategies for
  wireless sensor networks,'' {\em Proceedings of IEEE}, vol.~17, 2010.

\bibitem{jayakrishnan}
J.~Unnikrishnan and V.V.Veeravalli, ``Cooperative sensing for primary detection
  in cognitive radio,'' {\em IEEE Journal of selected topics in Signal
  Processing}, vol.~2, pp.~18--27, Feb 2008.

\bibitem{banavar}
S.~Dasarathan and C.~Tepedelenlioglu, ``Distributed estimation and detection
  with bounded transmissions over gaussian multiple access channels,'' {\em
  IEEE Trans. on Signal Processing}, vol.~62, no.~13, pp.~3454--3463, July
  2014.

\bibitem{no_noiseFC}
S.~Maleki, A.~Pandharipande, and G.~Leus, ``Energy-efficient distributed
  spectrum sensing for cognitive sensor networks,'' {\em IEEE Sensors Journal},
  vol.~11, no.~3, pp.~565--573, March 2011.

\bibitem{multipath}
N.~Reisi, S.~Gazor, and M.~Ahmadian, ``Distributed cooperative spectrum sensing
  in mixture of large and small scale fading channels,'' {\em IEEE Trans. on
  Wireless Communications}, vol.~12, no.~11, pp.~5406--5412, Nov 2013.

\bibitem{simon_alouni}
M.~K. Simon and M.-S. Alouini, {\em Digital Communication over Fading
  Channels}.
\newblock 2nd edition, John Wiley and Sons, 2005.

\bibitem{K_distri}
A.~Abdi and M.~Kaveh, ``K distribution: an appropriate substitute for
  rayleigh-lognormal distribution in fading-shadowing wireless channels,'' {\em
  IET Electronics Letters}, vol.~34, no.~9, pp.~851--852, 1998.

\bibitem{partial_coherent}
H.~Abdel-Ghaffar and S.~Pasupathy, ``Partially coherent, suboptimal detectors
  over rayleigh fading diversity channels,'' in {\em Communications Theory
  Mini-Conference Record, GLOBECOM}, pp.~76--80, 1994.

\bibitem{lehmann}
E.~Lehmann and J.~P. Romano, {\em Testing Statistical Hypotheses}.
\newblock 3rd edition, Springer, New York, 2005.

\bibitem{febi}
F.~Ibrahim and V.~Sharma, ``Novel distributed sequential nonparametric tests
  for spectrum sensing,'' in {\em Proceedings of IEEE GlobalSIP}, 2014.

\bibitem{sigman}
K.~Sigman, ``Appendix: A primer on heavy-tailed distributions,'' {\em Queueing
  systems}, vol.~33, pp.~261--275, 1999.

\bibitem{gut}
A.~Gut, {\em Stopped random walks : limit theorems and applications}.
\newblock Springer-Verlag, New York, 1988.

\bibitem{asmussen}
S.~Asmussen, {\em Applied Probability and Queues}.
\newblock 2nd edition, Springer, 2003.

\bibitem{asm_alb}
S.~Asmussen and H.~Albrecher, {\em Ruin Probabilities}.
\newblock 2nd edition, World Scientific, 2010.

\bibitem{foss}
S.G.Foss and S.Zachary, ``The maximum on a random time interval of a random
  walk with long tailed increments and negative drift,'' {\em Annals of applied
  probability}, vol.~1, pp.~37--57, 2003.

\bibitem{gaussiannonidentical}
H.~Barakat and Y.~Abdelkader, ``Computing the moments of order statistics from
  non-identical random variables,'' {\em Statistical Methods and Applications},
  vol.~13, pp.~15--26, 2004.

\bibitem{billingsley}
P.~Billingsley, {\em Probability and Measure}.
\newblock John Wiley and Sons, 1986.

\bibitem{zolotarev}
V.~M. Zolotarev, ``A one-sided interpretation and refinements of certain
  chebyshev-type inequality,'' {\em Selected translations in Math. Statistics
  and Probability}, vol.~12, 1973.

\bibitem{meiners}
A.~Iksanov and M.~Meiners, ``Exponential moments of first passage times and
  related quantities for random walks,'' {\em Electrical Communication Prob.},
  vol.~15, pp.~365--375, 2010.

\end{thebibliography}
\bibliographystyle{ieeetr}	% (uses file "ieeetr.bst")

\end{document}